\documentclass{revtex4}
\usepackage{amsmath, amssymb, bm, enumerate, graphicx, graphics, color,mathrsfs,hyperref}

\newcommand{\Lie}[0]{{\cal L}\, }

\newcommand{\tl}{\theta_{(\ell)}}
\newcommand{\tn}{\theta_{(n)}}

\newcommand{\ttau}[1]{\tilde{\tau}^{{#1}/3}}

\newcommand{\nn}{\nonumber}
\newcommand{\be}{\begin{equation}}
\newcommand{\ee}{\end{equation}}
\newcommand{\bea}{\begin{eqnarray}}
\newcommand{\eea}{\end{eqnarray}}

\newcommand{\tq}{\tilde{q}}

\newcommand{\A}{\textsf{A}}
\newcommand{\B}{\textsf{B}}
\newcommand{\C}{\textsf{C}}
\newcommand{\D}{\textsf{D}}

\newcommand{\ii}{\textsf{i}}
\newcommand{\jj}{\textsf{j}}
\newcommand{\kk}{\textsf{k}}

\newcommand{\tom}{\tilde{\omega}}

\newcommand{\vS}{\widetilde{\mbox{\boldmath$\epsilon$}}}

\newcommand{\norm}{| \mspace{-2mu} |}

\newcommand{\cV}{\mathcal{V}}

\begin{document}

\title{Spacetime near isolated and dynamical trapping horizons}
\date{\today}
\author {Ivan Booth}
\email{ibooth@mun.ca}
\affiliation{
Department of Mathematics and Statistics\\ Memorial University of Newfoundland \\  
St. John's, Newfoundland and Labrador, A1C 5S7, Canada \\
}

\begin{abstract}
We study the near-horizon spacetime for isolated and dynamical trapping horizons
(equivalently marginally outer trapped tubes). The metric is expanded relative to 
an ingoing Gaussian null coordinate and the terms of that expansion are explicitly 
calculated to second order. For the spacelike case, knowledge of the intrinsic
and extrinsic geometry of the (dynamical) horizon is sufficient to determine the 
near-horizon spacetime, while for the null case (an isolated horizon) more 
information is needed. In both cases spacetime is allowed to be of arbitrary dimension and the 
formalism accomodates both general relativity as well as more general field equations.

The formalism is demonstrated for two applications. 
First, spacetime is considered near an isolated horizon and the construction is both checked against the
Kerr-Newman solution and compared to the well-known near-horizon limit for stationary extremal black hole spacetimes. 
Second, spacetime is examined in the vicinity of a slowly evolving horizon and it is demonstrated that there is 
always an event horizon candidate in this region. The geometry and other properties of this null surface match 
those of the slowly evolving horizon to leading order and in this approximation the candidate evolves in a 
locally determined way. This generalizes known results for Vaidya as well as certain spacetimes known from 
studies of the fluid-gravity correspondence. 
\end{abstract}

\maketitle

\section{Introduction}
There is a large literature on geometrically defined black holes and their horizons. Trapped surfaces and apparent horizons were defined close to fifty 
years ago \cite{Penrose:1964wq}, but in the last two decades interest in both the mathematical and physical properties of these objects 
has increased.  Almost twenty years have passed since Hayward's original definition of 
trapping horizons\cite{hayward} while isolated and dynamical horizons have been studied for over a decade\cite{isoOriginal,ak}. 
In that time these geometric horizons have been widely studied in mathematical relativity 
(see for example \cite{badri, Dain:2011mv,Hollands:2011sy,Jaramillo:2011pg,Williams:2010pj, Nielsen:2010gm,Bengtsson:2010tj,
Williams:2009me,Bengtsson:2008jr, Andersson:2008up,vaidya, ExtremalPaper, bfbig, Galloway:2006ws,GourJara,billspaper, 
Gourgoulhon:2005ch,Andersson:2005gq,mttpaper,Ashtekar:2005ez,Ashtekar:2005cj,Schnetter:2005ea,Hayward:2005gi,
bendov,prl,isoGeom, isoMech,lewandowski,Liko:2007th,Liko:2007vi,Liko:2007mu,Liko:2008rr,Liko:2009qy})
but have also found applications in numerical relativity (examples include 
\cite{Nielsen:2010wq,Jaramillo:2011re,Jaramillo:2011rf,chu,Jasiulek:2009zf,Vasset:2009pf,Pollney:2007ss,Krishnan:2007pu,
Campanelli:2006fy,Schnetter:2006yt,gourgoul})
loop quantum gravity (for example \cite{Domagala:2004jt,Ashtekar:2004nd,Ashtekar:2000eq}), and  the fluid-gravity duality regime of the  
AdS-CFT correspondence\cite{fg3,fg2,fg1,Mukund}.

Very recently work has begun to study spacetime close to the horizon\cite{SEHreview,badri}. This paper presents the details of results first announced in \cite{SEHreview},
developing the necessary mathematics for stepping off of the horizon and studying the near-horizon spacetime. 
We expect this work to find many uses, but here we concentrate on the mathematical formalism followed by just two applications.
In the first we construct the spacetime around an extremal isolated horizon and show that at leading order it takes the familiar near-horizon 
form found in works such as \cite{hariOld,hariClass}. In the second we construct spacetime around a general slowly evolving horizon and
demonstrate the existence of a null surface that hugs the horizon. This candidate event horizon has previously been found for specific
spacetimes including Vaidya\cite{Mukund,vaidya,Nielsen:2010gm} and several black-brane spacetimes that show up in the fluid-gravity correspondence\cite{Mukund,fg1,fg2,fg3}. 

The core of our construction is a series expansion of the near-horizon metric: Eqn.~(\ref{metric_expansionII}). The metric is expressed 
in terms of horizon-based, ingoing Gaussian null coordinates and expanded relative to the ingoing (radial) affine parameter. This 
expression is universal and applies to spacetimes of arbitrary dimension with horizons of arbitrary signature. However, in the case of a
dynamical (spacelike) horizon, it  may be thought of as a generalization of the standard $(3\!+\!1)$-formulation of general relativity in which 
the Einstein equations are decomposed into constraints on the intrinsic and extrinsic geometry of a three-surface, along with 
evolution equations that determine how that geometry changes as the surface is propagated forwards in time. Similarly we will derive a 
set of constraint equations for the allowed geometry of a horizon along with evolution equations that describe how that geometry changes 
as one moves away from the horizon. These are used to construct spacetime in a neighbourhood of the horizon.


In a little more detail, we work with $(n\!+\!1)$-dimensional spacetimes and consider $n$-dimensional hypersurfaces which can be foliated into 
$(n\!-\!1)$-dimensional spacelike surfaces. This mirrors the structure of dynamical trapping horizons which come with a unique 
foliation\cite{Ashtekar:2005ez} as well as the marginally outer trapped tubes  found in numerical relativity  (which are 
built from $(n\!-\!1)$-dimensional  apparent horizons found on successive time slices). How the geometry of the $(n\!-\!1)$-leaves changes under deformations was studied in detail in
\cite{bfbig} and we import many results from that paper. However, that work must be supplemented in order to understand the full
geometry of horizons and other nearby $n$-dimensional surfaces. We also need to understand how the slices fit together and how that
fit changes under deformations. 

After studying the geometry of deformations in some generality we specialize to find the near-horizon spacetime metric in
Gaussian null coordinates constructed off of the horizon. Given a horizon $H$ and its foliation $S_v$, we take the (inward) null 
normals to the $S_v$ and consider the null geodesics that are tangent to those normals at $H$. Coordinates are then constructed 
taking the affine parameter $\rho$ along the geodesics as the off-horizon coordinate and Lie-dragging the on-horizon coordinates 
to the level surfaces of $\rho$. The components of the metric in these coordinates can be calculated (perturbatively) by considering 
how the geometry changes if $H$ is deformed
by the $\partial/\partial \rho$ vector field. The result is a series expansion of the metric where the individual terms are determined 
by quantities defined on the horizon. We explicitly calculate the terms of the series to second order. 

For a dynamical (spacelike) horizon in a vacuum spacetime, the intrinsic and extrinsic geometry of the horizon are sufficient to determine
those terms. In this case the construction is essentially equivalent to the standard $(3\!+\!1)$ initial value formulation of general relativity. 
For non-vacuum spacetimes one must also have information about the matter fields but the basic principle is unchanged. A spacelike
surface has a non-trivial future domain of dependence;  spacetime in that domain is determined by initial data on the surface. However, 
for an isolated (null) horizon things are different. In that case the formal series expansion is unchanged, however the terms of the 
series cannot be determined solely by standard initial data. The future domain of dependence of a null surface is empty and this is manifested
in our expansion by the fact that extra information beyond the basic horizon geometry is required to evaluate the terms. 

Apart from the application to extremal horizons, we are chiefly interested in spacetime near dynamical horizons. The null 
case is treated in detail in the programme recently initiated by Krishnan \cite{badri}. As in our approach he uses Gaussian
normal coordinates and rebuilds spacetime metrics (to second order) from the deformations of quantities such as the expansions and 
shears. However, his focus is isolated horizons and so he carefully treats the characteristic initial value problem, 
explaining what data must be specified (and where) in order to fully determine spacetime near such a null surface. 
%
%
Which of the formalisms is more useful will depend on the details of a  particular application.

The paper is organized as follows.  In Section \ref{Sigma} we briefly review the $(n\!+\!1)$-formulation of general relativity. The intent of this section is mainly to 
demonstrate the processes that will be used in later sections but in a setting that is more familiar to most readers. 
Section \ref{MainCalc} reviews the geometry of $(n\!-\!1)$-dimensional surfaces and the $n$-dimensional hypersurfaces 
that can be built from them and then applies that work to perturbatively reconstruct spacetime near such  hypersurfaces. 
Next, Section \ref{horizons} reviews definitions of the various types of geometric and causal horizons.
Section \ref{Extremal} applies the definition of an isolated horizon to reconstruct the spacetime around an
extremal isolated horizon while Section \ref{EventHorizon} works from slowly evolving horizons to demonstrate the existence of a horizon-hugging 
event horizon candidate.  As a cross-check it also compares this general result with analogous ones known for particular 
spacetimes. Section \ref{Discuss} summarizes this work and considers some future applications. Finally, working with Kerr-Newman 
spacetimes, Appendix
\ref{KerrEx} demonstrates both the construction of Gaussian null coordinates around an isolated horizon as well as how the 
near-horizon metric may be reconstructed from data specified on the horizon.


\subsection*{Notation}
Throughout we assume an $(n\!+\!1)$-dimensional spacetime $(M, g_{ab}, \nabla_a)$ and study embedded surfaces of dimensions $n$ and $(n\!-\!1)$. 
We use lower-case early-alphabet latin letters $\{a, b, c, \dots, g\}$ as abstract indices on the full spacetime but switch to 
greek letters $\{\alpha, \beta, \gamma \dots  \}$  when working with a coordinate chart. Similarly lower-case mid-alphabet latin letters 
$\{h, i , j, \dots p\}$ are used as abstract indices for tensors in $n$-dimensional surfaces while sans-serif versions of the same letters
$\{\mathsf{h}, \mathsf{i} ,\mathsf{j}, \dots \mathsf{p}\}$ are used for coordinates and tensor components relative to coordinates. 
Finally upper-case latin letters $\{A, B, C \dots \}$ are used as abstract indices on $(n\!-\!1)$-dimensional surfaces while their sans-serif versions
$\{\A, \B, \C \dots \}$ are used to indicate quantities written in terms of coordinates. 

The pull-back operator between surfaces will always be written as an $e$ with indices indicating which spaces it operates between. Thus the induced
metrics on $n$- and $(n\!-\!1)$-dimensional surfaces are respectively 
\be
q_{ij} = e_i^a e_j^b g_{ab} \; \; \mbox{and} \; \; \tq_{AB} = e_A^a e_B^b g_{ab}  \, , 
\ee
while tangent vectors to those same surfaces would push-forward to $TM$ via
\be
N^a = e_i^a V^i \; \; \mbox{and} \; \; \tilde{V}^a = e_A^a \tilde{V}^A \, . 
\ee
If we switch to coordinate charts so that (sections of) $n$- and $(n\!-\!1)$-dimensional surfaces are parameterized by functions
\be
x^\alpha = \mathcal{X}^\alpha (y^\mathsf{i})  \; \; \mbox{and} \; \; x^\alpha = \mathcal{Z}^\alpha (\theta^\A) \, , 
\ee
then 
\be
e_\mathsf{i}^\alpha = \frac{\partial \mathcal{X}^\alpha}{\partial y^\mathsf{i}} \; \; \mbox{and} \; \; 
e_\mathsf{A}^\alpha = \frac{\partial \mathcal{Z}^\alpha}{\partial \theta^\mathsf{A}} \, . 
\ee 
Fixing $\mathsf{i}$ or $\mathsf{A}$, these are also the components of the coordinate tangent vectors 
$\partial/\partial y^\mathsf{i}$ and $\partial/\partial \theta^\A$ (after they have been pushed-forward to $TM$). 

We follow the sign conventions of Wald \cite{wald} for such things as Riemann and extrinsic curvatures.

\section{Geometry of  spacelike hypersurfaces}
\label{Sigma}

To establish basic ideas about hypersurfaces and how their geometry changes under deformations we begin with a quick 
review of the $(n\!+\!1)$-formulation of general relativity. For more details see standard texts such as \cite{wald, eric, thomas}.
Analogous deformation calculations will be extensively used in subsequent sections and in the case of a spacelike
horizon, there will even be a similar horizon-based initial value formulation for the near-horizon spacetime.

\subsection{Basic geometry of an $n$-dimensional spacelike hypersurface}

Let $(\Sigma, q_{ij}, D_i)$ be a spacelike $n$-dimensional surface embedded in an $(n\!+\!1)$-dimensional spacetime $(M,g_{ab}, \nabla_a)$. 
As noted in the preamble, the induced metric on $\Sigma$ is 
\be
q_{ij} = e_i^a e_j^b g_{ab}  \, , 
\ee
while  the corresponding extrinsic curvature is
\be
K_{ij} = e_i^a e_j^b \nabla_a \hat{\tau}_b \, ,
\ee
where $\hat{\tau}^a$ is the future-oriented unit normal to $\Sigma$.

As for the elementary differential geometry of surfaces in Euclidean $\mathbb{R}^3$, the intrinsic metric and extrinsic curvature are not independent
but instead are related to each other as well as the curvature of the ambient spacetime through the Gauss-Codazzi equations. From the Gauss
relation one can show that
\be
G_{ab}  \hat{\tau}^a \hat{\tau}^b = \frac{1}{2} \left( R_\Sigma  + K^2 - K_{ij}K^{ij} \right)  \, , \label{preHamCon}
\ee
while from the Codazzi relation
\be
 e_i^a G_{ab} \hat{\tau}^b = D_j K_i^{\phantom{i} j}  - D_i K \label{preMomCon} \, .
\ee
In these equations, $R_\Sigma$ is the Ricci scalar for $\Sigma$, $K = q^{ij} K_{ij}$ is the trace of its extrinsic curvature 
and $G_{ab} = \mathcal{R}_{ab} - \frac{1}{2} \mathcal{R} g_{ab}$ is the Einstein tensor for $M$. Applying the Einstein equations 
\be
G_{ab} + \Lambda g_{ab} = 8 \pi T_{ab} 
\ee
to these, the Einstein tensor is replaced by terms involving the stress-energy tensor and cosmological constant. 
Then equations (\ref{preHamCon}) and (\ref{preMomCon}) respectively become the
\emph{Hamiltonian} and \emph{momentum} constraint equations. These conditions necessarily hold if $\Sigma$ is surface 
embedded in a full solution of the Einstein equations. 
 In particular, if  $(\Sigma, q_{ij}, K_{ij})$ is viewed as in instantaneous configuration that will be time-evolved into a full spacetime 
(the viewpoint taken in numerical relativity) then the fields must satisfy these constraints to be valid initial data.

%

\subsection{Deforming an $n$-dimensional spacelike hypersurface}
\label{SpaceDef}

\begin{figure*}
\includegraphics{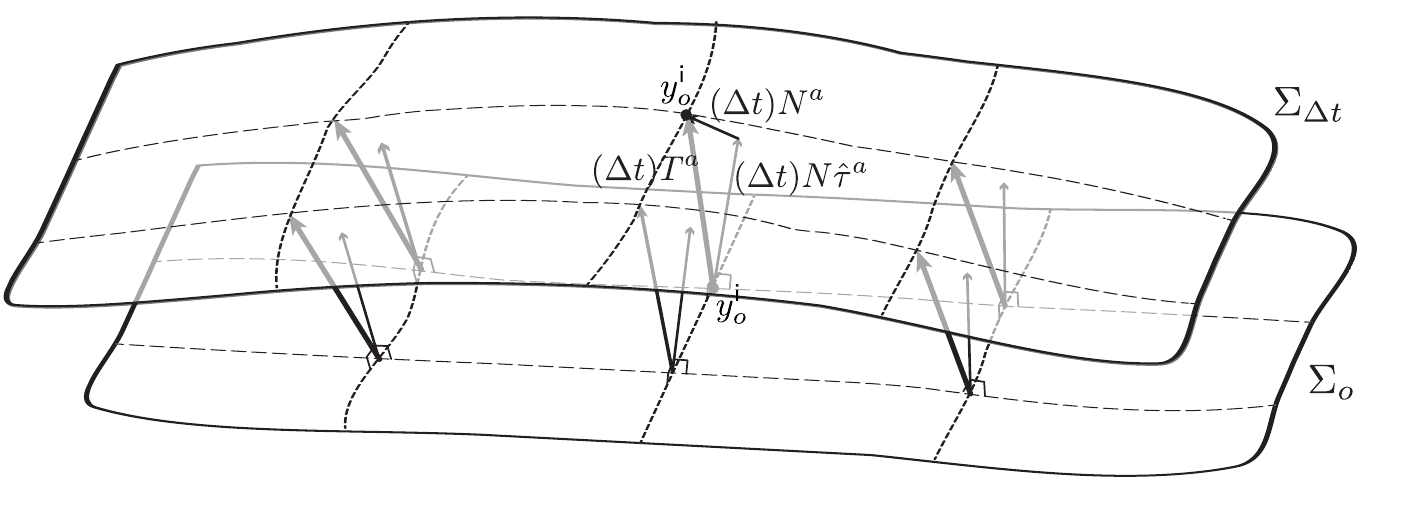}
\caption{Time evolution of a spacelike slice. The time-evolution vector $T^a$ deforms an initial surface $\Sigma_o$ into $\Sigma_{\Delta t}$. It can be
decomposed its parts perpendicular and parallel to $\Sigma$, hence defining a lapse function $N$ and shift-vector field $N^a = e^a_i N^i$. }
\label{SpacelikeDef}
\end{figure*}


Next, we consider how the induced metric and extrinsic curvature change if a hypersurface $\Sigma_o$ is deformed. 
This is easiest to understand if we switch to working 
with coordinate charts. Let  $\{x^\alpha \}$ be a set of $(n\!+\!1)$ coordinates over (at least some region of) $M$ and $\{y^\mathsf{i}\}$ be a set of
$n$ coordinates so that an initial (portion of) $\Sigma_o$ is parameterized by equations
\be
x^\alpha = \mathcal{X}^\alpha (y^\mathsf{i}) \, . 
\ee
Now a deformation may be defined by specifying a scalar field $N$ and vector field $N^i$ over $\Sigma_o$. These are respectively called the lapse and 
shift and used to construct an $n$-dimensional vector field
\be
T^a = N \hat{\tau}^a + N^a = N \hat{\tau}^a + e^a_i N^i \, . 
\ee
In turn this is used to (infinitesimally) generate a new surface $\Sigma_{\Delta t}$ defined by 
\be
\mathcal{X}^\alpha (y^\mathsf{i}) \rightarrow \mathcal{X}^\alpha (y^\mathsf{i}) + (\Delta t) T^\alpha (y^\mathsf{i}) \, . 
\ee
The mapping also identifies points on the hypersurfaces (essentially by Lie-dragging coordinates between surfaces) as shown in FIG.~\ref{SpacelikeDef}.
Derivatives relative to $T^a$ are defined in terms of the usual limits. For example the deformation of the induced metric is
\begin{equation}
\delta_T {q}_{\mathsf{i} \mathsf{j}} = \lim_{\Delta t \rightarrow 0} \frac{\left.q_{\mathsf{i} \mathsf{j}} \right|_{\mathcal{X}+(\Delta t) T} - \left.q_{\mathsf{i} 
\mathsf{j}} \right|_{\mathcal{X}} }{\Delta t} \, , \label{deltaTq}
\end{equation}
where the metrics on the two surfaces may be directly compared since they are expressed relative to the same (Lie-dragged) coordinate system.
This is  a covariant construction.

Of course one doesn't use the limit definition to actually compute deformations. Instead calculations are based on the understanding that the 
deformations effectively extend the coordinate system $\{ y^\mathsf{i} \}$ off of $\Sigma_o$ and supplement it with another coordinate $t$ 
such that
\be
T =\frac{\partial}{\partial t}  \; \; \mbox{and} \; \; e_\mathsf{i} = \frac{\partial}{\partial y^\mathsf{i}} \, .
\ee
These are coordinate vectors and so their Lie brackets vanish. In particular
\be
\Lie_T e_i^\alpha = 0   \label{LTe}
\ee
and in turn if follows that
\be
e_i^a \Lie_T \hat{\tau}_a =  0 \; \; \Longrightarrow \; \;  e^a_i T^b \nabla_b \hat{\tau}_a =  D_i  N + K_{ij} N^j\, .  \label{LTeCons}
\ee

Deformations are calculated by taking Lie derivatives with respect to $T^a$ with appropriate applications of (\ref{LTeCons}) to 
convert derivatives normal to $\Sigma$ into derivatives tangent to it. The
derivative of the induced metric can be calculated without this condition 
\be
\delta_T q_{ij} = e_i^a e_j^b \Lie_T g_{ab} = 2 N K_{ij} + \Lie_N q_{ij}  \label{Ltq}
\ee
however the deformation of $K_{ij}$ is a little more complicated. Starting from  
\be
\delta_T K_{ij} = e_i^a e_j^b \Lie_T (\nabla_a \hat{\tau}_b ) 
\ee
a certain amount of algebra along with an application of the Gauss relation gives:
\bea
\delta_T K_{ij} =  D_i D_j N - N (R_{ij} - 2 K_{ik} K_j^{\phantom{j}k} + K K_{ij}) +  \Lie_N K_{ij} + \left(e_i^a e_j^b - \frac{1}{2} q_{ij} g^{ab} \right) G_{ab} \, , 
\label{LTK} 
\eea
where $R_{ij}$ is the $n$-dimensional Ricci tensor for the surface. 

Of course, the best known example of a deformation is the time evolution of initial data in the $(3\!+\!1)$-formulation of 
general relativity. Given an initial surface $\Sigma$, the Einstein equations are equivalent to the constraints (\ref{preHamCon}) and 
(\ref{preMomCon}) along with the time evolution of the extrinsic curvature (\ref{LTK}). Though the details will differ, this is the 
perspective that we will adopt when studying the spacetime around a dynamical horizon -- we will take its geometric specification 
as initial data and then evolve that data to construct spacetime in a neighbourhood of that horizon. 

\subsection{Spacetime near a spacelike hypersurface}
\label{SpaceNearSigma}

We can also use this knowledge of deformations to perturbatively construct spacetime close to a spacelike surface. 
Working in geodesic normal coordinates based on  $\Sigma_o$ the spacetime metric takes the form
\be
ds^2 = - d \tau^2 + {h}_{\mathsf{i} \mathsf{j}} dy^\mathsf{i} dy^\mathsf{j}  \,  , 
\ee
where $\tau$ measures proper time along the geodesics and the $h_{\mathsf{i} \mathsf{j}}$ are the components of the spacelike $n$-dimensional metric 
on surfaces $\Sigma_\tau$ of constant $\tau$. It is straight-forward to expand this as a Taylor series (in $\tau)$ around the initial data
on $\Sigma_o$. To second order we have
\bea
ds^2 \approx -d \tau^2 +  \left( q_{\mathsf{i} \mathsf{j}}  + \tau  q'_{\mathsf{i} \mathsf{j}}  + \frac{\tau^2}{2} q''_{\mathsf{i} \mathsf{j}} \right) dy^\mathsf{i} dy^\mathsf{j} \, ,
\eea 
where $q_{\ii \jj}$ is the induced metric on $\Sigma_o$, 
\be
q'_ {\mathsf{i} \mathsf{j} } = \delta_{\hat{\tau}} q_ {\mathsf{i} \mathsf{j} } = 2 K_ {\mathsf{i} \mathsf{j} }
\ee
and 
\be
q''_{\mathsf{i} \mathsf{j} } = \delta_{\hat{\tau}} \left( \delta_{\hat{\tau}}q_ {\mathsf{i} \mathsf{j} }  \right) = 
 -  2 (R_{\ii \jj} - 2 K_{\ii\kk} K_\jj^{\phantom{\jj}\kk} + K K_{\ii\jj})  +2 \left(e_\ii^\alpha e_\jj^\beta - \frac{1}{2} q_{\ii\jj} g^{\alpha \beta} \right) G_{\alpha \beta} \, , 
\ee
since geodesic coordinates have $N=1$ and vanishing shift vector. 

Initial data on a spacelike surface fully determines the spacetime in its future domain of dependence \cite{wald,thomas}. Our Taylor expansion is 
consistent with this result. Momentarily restricting attention to vacuum spacetimes that are solutions of the Einstein equations 
(so that $G_{\alpha \beta} = 0$), knowledge of $q_{ij}$ and $K_{ij}$ on $\Sigma_o$ allows us to calculate all orders of derivatives of $q_{\ii \jj}$.
The first derivative of induced metric is determined by the extrinsic curvature while the derivative of the extrinsic curvature is determined by the
induced metric and extrinsic curvature. This closes the circle and so, based on the initial data, we can construct the Taylor expansion to 
all orders. 

Though this construction works for all spacelike surfaces we will see in future sections that it is not well suited to the study of spacetime close
to a near-equilibrium dynamical horizon. Such surfaces are ``nearly-null'' and do not comfortably fit with constructions based on timelike
normals. Instead we will construct an analogous formalism based on Gaussian null coordinates. This will also allow us to use the same
language to talk about the spacetime near a null horizon (though in that case the specification of initial data is quite different).


\section{Geometry of hypersurfaces of arbitrary signature}
\label{MainCalc}

The previous section has reviewed several important ideas. First, a surface is characterized by its intrinsic and extrinsic geometry and in general 
these are not independent: they are linked to each other and the curvature tensor of the full spacetime through the Gauss-Codazzi equations. Second,
we can calculate how the surface geometry changes if the surface is deformed by a vector field: computationally this amounts to taking Lie derivatives
of surface quantities with the extra condition that the Lie derivative of pull-back operators vanishes. Third, given a coordinate system on the 
surface and a preferred vector field (in this case the tangent vector field to the congruence of timelike geodesics normal to $\Sigma$) we can 
construct a coordinate system in a neighbourhood of $\Sigma$ and use the deformation results to construct the spacetime metric relative to 
that coordinate system. The intrinsic metric and extrinsic curvature tensor of $\Sigma$ are good initial data (provided that they satisfy the constraints).
Via the deformation/time evolution equations they fully determine the vacuum spacetime metric close to the horizon.

These ideas form the foundation of the $(3\!+\!1)$ formulation of general relativity where an initial instantaneous configuration
is evolved into a full spacetime via evolution equations. Ultimately, as much as possible, we wish to mirror this construction where the 
initial data is a geometric horizon $H$: given its intrinsic and extrinsic geometry we would like to construct spacetime in a 
neighbourhood of $H$. Unfortunately the standard formalism is not sufficient for our purposes. The class of geometric horizons includes
the null (isolated) horizons and in that
case standard initial data on the horizon is not sufficient to reconstruct the spacetime \cite{badri,friedrich}. Further, even when they are 
spacelike we will often be interested
in the regime where they are 
``almost'' null. Thus we do not wish to base our formalism around a timelike unit normal vector $\hat{\tau}^a$: it is not well-defined for truly null 
surfaces and is inconvenient for ``almost null'' surfaces. 

The standard formalism is also not sufficient in that we wish to work with spacetimes of arbitrary dimensions as well as allow 
for generalizations of the Einstein equations. However these are relatively minor issues. Switching to higher dimensions 
is straightforward while the Einstein equations are not actually a fundamental part of the formalism. Most of it is general to any 
spacetime and the Einstein equations are only used to constrain the potential spacetimes under consideration. 
 
This section reformulates the standard formalism to accommodate our goals. As such we consider $n$-dimensional surfaces that 
can be defined as the smooth union of a set of spacelike $(n\!-\!1)$-dimensional surfaces $S_v$: $H = \{ \cup_v S_v \}$ for some range of the surface label $v$. In order to ensure maximum generality and  applicability, at this stage we do not impose any restrictions on the overall signature of $H$
and neither do we assume that it has any particular geometric properties (for example we do not assume that it is marginally outer trapped).

At this stage, our concern is expanding the near-horizon geometry in terms of quantities specified on the horizon. Later sections 
will consider how these quantities are (or are not) specified in an initial value formulation. 

%
%
 
%
  
 \subsection{Geometry of $(n\!-\!1)$-dimensional spacelike surfaces}
 We begin by reviewing the geometry of the $(n\!-\!1)$-dimensional building blocks of $H$. As in the previous section we split this into a consideration 
 of basic geometry followed by a study of deformations. This mathematics is well-known and has been derived and re-derived many times. 
 That said, our immediate  reference (in which many more details can be found) is \cite{bfbig}. 

 \subsubsection{Basic geometry of $(n\!-\!1)$-dimensional surfaces}
 \label{BasicGeometry}

Let $(S,\tq_{AB}, d_A)$ be an $(n\!-\!1)$-dimensional 
spacelike surface embedded in a time-orientable $(n\!+\!1)$-dimensional $(M,g_{ab}, \nabla_A)$.
The induced metric on  $S$ is the pull-back of the full spacetime metric:
\bea
\tq_{AB} = e_A^a e_B^b g_{ab}   
\eea 
and this determines the full intrinsic geometry of $S$ including the metric compatible covariant derivative $d_A$ and the Riemann tensor
$\tilde{R}_{ABCD}$. 
In the particularly important case where $S$ is two-dimensional ($n=3$) we have 
\be
\tilde{R}_{ABCD} = \frac{1}{2} \tilde{R} \left(\tq_{AC} \tq_{BD} - \tq_{AD} \tq_{BC} \right) \, , 
\ee
for the two-dimensional Ricci scalar $\tilde{R}$. 

The normal space to $S$ is two-dimensional. It is spanned by a pair of future-oriented null vectors $\ell^a$ 
and $n^a$ and we assume that properties of $S$ and the spacetime are such that these can respectively identified as outward and
inward pointing. The direction of these vectors is fixed, but their scaling isn't. We remove one degree of freedom by requiring that
they be cross-normalized so that 
\be
\ell \cdot n = -1
\ee
which leaves a single degree of rescaling freedom:
\be
\ell \rightarrow e^f \ell \; \; \; \mbox{and} \; \; \;  n \rightarrow e^{-f} n \, , \label{scaling}
\ee
for an arbitrary function $f$. However, independent of that particular choice of scaling we have
\begin{equation}
\tq^{ab} \equiv e^a_A e^b_B \tq^{AB} =  g^{ab} + \ell^a n^b + n^a \ell^b \label{tq} \, . 
\end{equation}
If $n=3$ then the induced area form on $S$ relative to that of the full spacetime is
\be
\tilde{\epsilon}_{AB} = e_A^a e_B^b (\epsilon_{abfg} \ell^f n^g) \, . 
\ee
Similarly for $n=4$ and $5$ the volume forms are
\be
\tilde{\epsilon}_{ABC} = e_A^a e_B^b e_C^c (\epsilon_{abcfg} \ell^f n^g) \; \; \mbox{and} \; \; 
\tilde{\epsilon}_{ABCD} = e_A^a e_B^b e_C^c e_D^d (\epsilon_{abcdfg} \ell^f n^g) \, . 
\ee
The generalization to even higher dimensions is obvious. 

Just as extrinsic geometry of $\Sigma$ was determined by how $\hat{\tau}_a$ varied over the surface, the extrinsic geometry of 
$S$ can be understood by considering how these null normals vary along the surface. 
We have the extrinsic curvature analogues:
\begin{equation}
k^{(\ell)}_{AB} = e_A^a e_B^b \nabla_a \ell_b \; \; \mbox{and} \; \; k^{(n)}_{AB} = \tq_A^a \tq_B^b \nabla_a n_b 
\end{equation}
as well as the connection on the normal bundle:
\begin{equation}
\tom_A = - e_A^a n_b \nabla_a \ell^b \, . 
\end{equation}
These measures of the intrinsic and extrinsic geometry are related to each other and the geometry of the ambient spacetime by analogues of the 
Gauss, Codazzi, and Ricci equations (see \cite{bfbig} for details). 

Note however that compared to the codimension-one spacelike case, there is an extra complication in dealing with these extrinsic curvature quantities.  
For spacelike codimension-one hypersurfaces, the extrinsic curvature is a uniquely defined geometric quantity. Here things are a little more complicated. Under the
rescalings defined by (\ref{scaling})
\begin{equation}
k^{(\ell)}_{AB} \rightarrow e^f k^{(\ell)}_{AB} \, , \;  \; k^{(n)}_{AB} \rightarrow e^{-f} k^{(n)}_{AB} \, \mbox{and} \; \; \tom_A \rightarrow \tom_A + d_A  f \,. 
\end{equation}
and so they have a gauge dependence. For now we accept this uncertainty but when constructing the spacetime around $H$  in Section 
\ref{n-geometry} we will fix the gauge. 

The traces of the extrinsic curvatures and their trace-free parts are important 
enough to have their own names. For a general element $X^a = \alpha \ell^a - \gamma n^a$ of the normal space we write
\be
k^{(X)}_{AB} = e_A^a e_B^b \nabla_a X_b = \frac{\theta_{(X)}}{(n\!-\!1)} \tq_{AB} + \sigma^{(X)}_{AB} \, ,
\ee
where the trace $\theta_{(X)} = \tq^{AB} k^{(X)}_{AB}$ is  the \emph{expansion} and the trace-free $\sigma^{(X)}_{AB}$ is the \emph{shear}. 

\subsubsection{Deforming an $(n\!-\!1)$ surface}


Just as we studied how the geometry of $\Sigma$ is changed by a deforming vector field we can also examine the deformations of $S$.  
The process is essentially the same. Parameterize $S$ in terms of a coordinate chart $\theta^\A$: 
$x^\alpha = \mathcal{X}^\alpha (\theta^\A)$.
Then,  for a deformation vector field $X^a$:
\begin{equation}
\mathcal{X}^\alpha (\theta^\A) \rightarrow \mathcal{X}^\alpha (\theta^\A) + \epsilon X^\alpha (\theta^\A)
\end{equation}
defines a new surface $S_\epsilon$ by deforming $S$ a coordinate distance $\epsilon$ in the direction $X^\alpha$. It identifies points
on the $(n\!-\!1)$-surfaces (essentially by Lie-dragging coordinates between surfaces) and so we have
\be
\Lie_{X} e_A^a  = 0 \, . 
\label{TanCond}
\ee
We can then examine how the geometry changes under these deformations. 
For our purposes, it will be sufficient to restrict attention to normal deformations so that 
\begin{equation}
X^a = \alpha \ell^a - \gamma n^a \, ,
\end{equation}
for some functions $\alpha$ and $\gamma$. 
Then 
it is straightforward to see that
\be
\delta_X \tq_{AB} = e_A^a e_B^b \Lie_X g_{ab} = 2 k^{(X)}_{AB}  = 2 (\alpha k^{(\ell)}_{AB} - \gamma k^{(n)}_{AB} ) \, , \label{dXq}
\ee
and (dropping the indices) 
\be
\delta_X \tilde{\epsilon} = \theta_{(X)} \tilde{\epsilon} = (\alpha \tl - \gamma \tn) \tilde{\epsilon} \label{dXep} \, .  
\ee
for the volume-form\footnote{Volume is used here in a general sense. For $n=2$ it would be a length, $n=3$ an area, $n=4$ a volume and 
$n>4$ an $(n-2)$-dimensional hypervolume.}. 
It is then obvious why we call the traces expansions.
They  tell us how the volume elements change while the shears are the part of the evolution that deforms $S$ 
(but does not change its volume).

%

As was the case for $n$-dimensional hypersurfaces, deformations of the extrinsic curvature quantities are more complicated. In addition to applications of
the Gauss, Codazzi and Ricci relations we again need to convert off-horizon derivatives into ones tangent to the $S$. As in the previous section the 
key to this is (\ref{TanCond}) and this time the analogues to (\ref{LTeCons}) are:
\begin{eqnarray}
X^b \nabla_b \ell_a & = & -d_a \gamma + \tom_a \gamma + \kappa_{X} \ell_a \label{XlXn} \; \; \mbox{and}\\
X^b \nabla_b n_a & = & d_a \alpha + \tom_a \alpha - \kappa_{X} n_a \nn \, ,
\end{eqnarray}
where $d_a f = e_a^A  d_A f$, $\tom_a = e_a^A \tom_A$, $\kappa_X = - X^a n_b \nabla_a \ell^b$ and the index-reversed 
$e_a^A = g_{ab} e^b_B \tq^{AB}$. Note that 
$\kappa_X$ is a gauge dependent quantity whose value depends on how the scaling of the null
vectors changes off the original surface. Under rescalings (\ref{scaling}) of the null vectors:
\be
\kappa_{X} \rightarrow \kappa_X + \Lie_X  f \, . 
\ee
In Section \ref{n-geometry} we will also gauge-fix this quantity but for now leave it undetermined. 
Because we have only considered normal deformations, there are no extrinsic curvature terms in (\ref{XlXn}) as compared to (\ref{LTeCons}).

For details of the deformation calculations see \cite{bfbig}, here we will just list results. 
First deforming the extrinsic curvature $k^{(\ell)}_{AB}$ we find that 
\bea
\delta_X k^{(\ell)}_{AB} & = &  \label{dkl}
  - d_A d_B \gamma + 2 \tom_{(A} d_{B)} \gamma 
  + \kappa_X k^{(\ell)}_{ AB} \\
&& +  \alpha \left(k^{(\ell)}_{AC} k^{(\ell) C}_B 
  - e_A^a \ell^b e_B^c \ell^d C_{abcd} 
  - \frac{1}{(n\!-\!1)} \tq_{AB} \mathcal{R}_{cd} \ell^c \ell^d \right) \nn\\
  && + \gamma \left( \frac{1}{2} \tilde{R}_{AB} + d_{(A} \tom_{B)} - \tom_A \tom_B
  + \frac{1}{2} \left[ \tl k^{(n)}_{AB} + \tn k^{(\ell)}_{AB} \right]  
  -  2 k^{(\ell)}_{C(A} k^{(n) C}_{B)} - \frac{1}{2} e_A^a e_B^b \mathcal{R}_{ab} 
   \right) \, , \nn 
\eea
where $C_{abcd}$  is the $(n\!+\!1)$-dimensional Weyl tensor and $\tilde{R}_{AB}$ is the 
$(n\!-\!1)$-dimensional Ricci tensor. In the usual way parentheses indicate a symmetrization of indices so, for example,
\be
\tom_{(A} d_{B)} \gamma = \frac{1}{2} \left(\tom_A d_B \gamma + \tom_B d_A \gamma \right) \, . 
\ee
The deformation of the extrinsic curvature $k^{(n)}_{ab}$ is
\bea
\delta_X k^{(n)}_{AB} & = & \label{dkn}
   d_A d_B \alpha + 2 \tom_{(A} d_{B)} \alpha 
  - \kappa_X k^{(n)}_{ AB} \\
&& -  \gamma \left(k^{(n)}_{AC} k^{(n) C}_B 
  - e_A^a n^b e_B^c n^d C_{abcd} 
  - \frac{1}{(n\!-\!1)} \tq_{AB} \mathcal{R}_{cd} n^c n^d \right) \nn\\
 && - \alpha \left( \frac{1}{2} \tilde{R}_{AB}  - d_{(A} \tom_{B)} - \tom_A \tom_B
  + \frac{1}{2} \left[ \tn k^{(\ell)}_{AB} + \tl k^{(n)}_{AB} \right]  
 -  2 k^{(n)}_{C(A} k^{(\ell) C}_{B)}  -  \frac{1}{2} e_A^a e_B^b \mathcal{R}_{ab} 
  \right) 
\, \, . \nn
\eea
Given (\ref{dkl}) one can easily find (\ref{dkn}) by interchanging $\ell$ and $n$ and then sending $\alpha \rightarrow - \gamma$
and $\gamma \rightarrow - \alpha$ (which together also mean $\tom_a \rightarrow - \tom_a$ and $\kappa_X \rightarrow - \kappa_X$). 

As noted, null expansions are particularly important in discussions of geometric horizons and so we also list the deformations
of the traces of the extrinsic curvatures:
\begin{eqnarray} \label{explderiv}
   \delta_X \theta_{(\ell)}&=& \phantom{-} \kappa_X \tl 
   - {d}^{\, 2} \mspace{-2mu} \gamma  
   + 2 \tilde{\omega}^{A} {d}_{A} \gamma \\ 
   &&  + \gamma \left[ \frac{\tilde{R}}{2} +  {d}_{A} \tilde{\omega}^{A}  - \norm \tilde{\omega} \norm^2
     - \frac{1}{2} \mathcal{R}_{ab} \tq^{ab} + \tl \tn \right] \nn 
      - \alpha \left[ \norm\sigma_{(\ell)}\norm^{2} 
     + \mathcal{R}_{ab} \ell^{a} \ell^{b}
     + \frac{1}{(n\!-\!1)} \theta_{(\ell)}^{2} \right]  \, \, , 
\eea
and
\bea  \label{expnderiv}
  \delta_X \theta_{(n)}  &=& - \kappa_X \tn +
  {d}^{\, 2} \mspace{-2mu} \alpha 
    + 2 \tilde{\omega}^{A} {d}_{A} \alpha \\
&&   - \alpha \left[  \frac{\tilde{R}}{2} -  {d}_{A} \tilde{\omega}^{A}  - \norm \tilde{\omega} \norm^2
     - \frac{1}{2} \mathcal{R}_{ab} \tq^{ab} + \tl \tn \right] \nn
  + \gamma \left[ \norm\sigma_{(n)}\norm^{2} 
    + \mathcal{R}_{ab} n^{a} n^{b} 
    + \frac{1}{(n\!-\!1)}\theta_{(n)}^{2} \right] \,  
      \, .
\eea
where $d^2 = d^A d_A$, $\norm \tom \norm^2 = \tom^A \tom_A$,  $\tilde{R}$ is the $(n\!-\!1)$-dimensional Ricci scalar for $S_v$,
$\norm \sigma_{(\ell)} \norm^2 = \sigma_{(\ell)}^{AB} \sigma^{(\ell)}_{AB} $ and $\norm \sigma_{(n)} \norm^2 = \sigma_{(n)}^{AB} \sigma^{(n)}_{AB}$.

Finally the variation of the connection one-form is
\begin{eqnarray}  
\delta_X \tilde{\omega}_A & = & - \theta_{(X)} \tom_A + d_A \kappa_X - d^B k_{AB}^{(X_{\!\perp}\!)} + \alpha d_A \tl + \gamma d_A \tn + e_A^a \mathcal{R}_{ab} X_{\!\perp}^b \, . \label{dtom}
\end{eqnarray}
Here $X_{\! \perp}^A \equiv \alpha \ell^a + \gamma n^a$ is normal to $X^a$ while $\theta_{X}$ and $k_{AB}^{(X_{\!\perp}\!)}$ are defined in the obvious 
way. If consulting \cite{bfbig} for details of this particular calculation, note that the last line equation (2.26) of that reference misses an overall factor 
of $1/2$.

\subsection{Building an $n$-dimensional hypersurface from $(n\!-\!1)$-surfaces}
\label{GenSurf}
%
%
%
%
\label{n-geometry}
Next, we put these $(n\!-\!1)$-dimensional pieces together into an $n$-dimensional surface. As is now our standard procedure, we start by considering 
the intrinsic and extrinsic geometry and then move on to consider deformations of that geometry.

\subsubsection{Intrinsic geometry}

\begin{figure*} 
\includegraphics{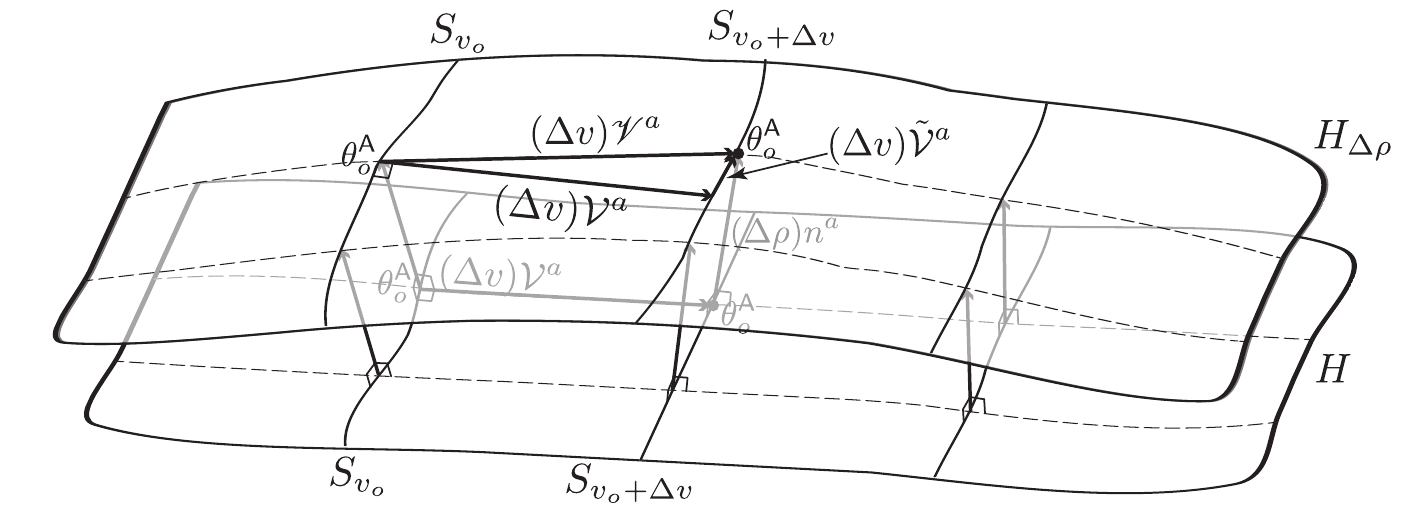}
\caption{This figure is similar in appearance to FIG.~\ref{SpacelikeDef} and represents a similar situation. The bottom sheet $H$ is foliated by 
$(n\!-\!1)$-dimensional surfaces $S_v$ (drawn as solid lines). On $H$, $\mathscr{V}^a$ is the evolution vector field evolving surfaces into each other 
from the left-hand to the right-hand sides and identifying points on those surfaces. 
The null normal $n^a$ is used to deform $H$ into $H_{\Delta \rho}$. The evolution vector field $\mathscr{V}^a$ on $H_{\Delta \rho}$ will usually 
no longer be normal to the $S_v$ after the translation. }
\label{DefH}
\end{figure*}

Let $H$ be an $n$-dimensional hypersurface which is the union of a set of spacelike $(n\!-\!1)$-dimensional surfaces $S_v$ with  $v$ labelling
the surfaces. Regardless of the signature of $H$, we can define a unique vector field $\mathcal{V}^i$ on $H$ that is 
1) normal to the $S_v$, 2) tangent to $H$ and 3) satisfies $\mathcal{L}_{\cV} v = 1$. Then a general \emph{evolution vector field} $\mathscr{V}^i$
that evolves leaves of the foliation into each other may be written as
\be
\mathscr{V}^i = \mathcal{V}^i + \tilde{\mathcal{V}}^i \, , 
\ee
where $\tilde{\mathcal{V}}^i$ is tangent to the $S_v$.
The canonical example of an evolution vector field arises if we impose a foliation-compatible coordinate system $z^\textsf{i} = \{ v, \theta^\A \}$ on $H$. 
Then
\be
\mathscr{V} = \frac{\partial}{\partial v} 
\ee
is an evolution vector field. For such a choice we have $\Lie_\mathscr{V} \theta^\A = 0$ and so it follows that 
\be
 \tilde{\mathcal{V}}^\A  = - \Lie_\cV \theta^\A \, . 
\ee
In analogy with the evolution vector field in the $(n\!+\!1)$-formulation we will refer to $\tilde{\cV}^A$ as a shift vector field.

Regardless of the value of the shift vector, the foliation parameter can be used to fix the scaling of the null vectors. We scale the null vectors so that
\be
\mathcal{V}^a = \ell^a - C n^a \, , 
\ee
for some \emph{expansion parameter} $C$\footnote{By making this choice we do not allow for situations where $H$ becomes parallel to $n$. 
For apparent/trapping this can happen under certain extreme situations where the horizon ``jumps'' \cite{bendov, mttpaper}. However in this paper where we focus on 
near-equilibrium horizons, we can safely disregard this regime. }. Note that if $C>0$, $H$ is spacelike, while if $C< 0$ it is timelike and if $C=0$ it is null. 
Given this construction, the scaling freedom of the null vectors is restricted to the freedom to reparameterize the foliation labelling. 
For an alternative labelling $\tilde{v} = \tilde{v}(v)$ we would have
\be
[d \tilde{v}]_a = \frac{1}{\alpha(v)} [d v]_a  \; \Rightarrow \; \widetilde{\mathcal{V}}^a = \alpha(v) \mathcal{V}^a
\ee
where $\alpha(v) = \frac{d v}{d\tilde{v}}$ is constant over each $S_v$. Then 
\be
\tilde{\ell}^a = \alpha  \ell^a  \, ,  \;  \tilde{n}^a = \frac{1}{\alpha} n^a \;  \; \mbox{and} \;  \tilde{C} = \alpha^2 C \, . \label{ReLabel}
\ee
Under this restricted class of rescalings (with $\alpha$ constant over each individual $S_v$), $\tom_a$ is invariant. 

Next consider the intrinsic and extrinsic geometry of the full $H$. Relative to  $\{z^{\textsf{i}} = (v , \theta^{\textsf{A}}) \}$ the intrinsic metric is 
\be
d\Sigma^2 = q_{\textsf{ij}} dz^i dz^j  = (2 C+\tq_{AB} \tilde{\cV}^A \tilde{\cV}^B) dv^2 + 2 \tq_{\A \B} \tilde{\cV}^\A dv d \theta^\B+  \tq_{\textsf{AB}} d \theta^{\textsf{A}} d\theta^{\textsf{B}} \, , \label{q0}
\ee
where $\tq_{\textsf{AB}}$ is, as usual, the induced metric on the $(n\!-\!1)$-surfaces. Note that this coordinate form explicitly 
demonstrates how the sign of $C$ determines the signature of $H$. As would be expected, the inverse metric $q^{ij}$ is not well-defined if 
$C=0$ (that is when $H$ is null). 
  
Shifting our attention to  the extrinsic geometry, a future-oriented normal one-form to $H$ is given by
\be
\tau_a = \ell_a + C n_a \, 
\ee
and so straightforward calculations demonstrate that the associated extrinsic curvature (again in coordinate form) is
\bea
 K^{(\tau)}_{\textsf{ij}} dz^\ii dz^\jj & \equiv & (e_\ii^\gamma e_\jj^\delta \nabla_\gamma \tau_\delta)  dz^\ii dz^\jj  \label{K0} \\
 &=& \left( 2 C {\kappa}_{\cV} - \Lie_{\cV} C \right) dv^2 + \left(k^{(\ell)}_{\A \B} 
 + C k^{(n)}_{\A \B} \right) d \theta^\A d \theta^\B    + \left( 2 C \tom_\A - d_\A C \right) (dv d \theta^\A + d \theta^A dv) \nn\, ,
\eea
where 
\be
\kappa_{\cV} = - \cV^a n_b \nabla_a \ell^b
\ee
is again the gauge-dependent quantity that measures how the scaling of the null vectors changes as one moves between $S_v$. 
Since we wish to allow for all values of $C$, we have not unit-normalized $\tau_a$.  
If  $H$ is spacelike, the usual extrinsic curvature of Section \ref{Sigma}, defined relative to the timelike unit normal $\hat{\tau}_a$, is
\be
K^{(\hat{\tau})}_{ij} = \frac{1}{\sqrt{2C}} K^{(\tau)}_{ij} \, . 
\ee

Thus, it is clear that the intrinsic and extrinsic geometry of $H$ is fully specified by the set of fields:
\be
( C, \tilde{\cV}^A, \tq_{AB},  \kappa_{\cV}, \tom_A, k^{(\ell)}_{AB}, k^{(n)}_{AB}) \, ,
\ee
and so, at least if $H$ is spacelike, we would expect to be able to use these as initial data for evolution into a full spacetime. That said, 
just as $q_{ij}$ and $K_{ij}$ are related to each other and the ambient geometry by constraints, our new fields are
also not all independent.  Most obviously, from (\ref{dXq}) we know that
\be
\Lie_{\cV} \tq_{AB} =2( k^{(\ell)}_{AB} - C k^{(n)}_{AB})  \, . 
\ee
Thus $k^{(\ell)}_{ab}$ can always be found from $C$, $k_{AB}^{(n)}$ and $\tq_{AB}$ and doesn't need to be independently specified. 

The other relations should be equivalent to the Hamiltonian and momentum constraints: the easiest way to identify how is to match equations by 
the components of the Einstein tensor that appear in them. Doing this we find that (\ref{explderiv})$-C\times$(\ref{expnderiv}) gives us
\bea
G_{ab} \tau^a \tau^b & = & \kappa_\cV \theta_{(\tau)} - \Lie_\cV \theta_{(\cV)}  - \tn \Lie_{\cV} C - d^2 C + 2 \tom^A d_A C
 + C \left( \tilde{R} - 2 \| \tom \|^2 + 2 \tl \tn \right)   \label{TVt}\\
& & - \left( \| \sigma_{(\ell)} \|^2 + C^2 \| \sigma_{(n)} \|^2 + \frac{1}{(n\!-\!1)} \left[\tl^2 +C^2 \tn^2 \right]  \right) \nn
\eea
which for spacelike $H$ is equivalent to the Hamiltonian constraint. The momentum constraint comes in two pieces. 
First  (\ref{explderiv})$+C\times$(\ref{expnderiv}) gives us 
\bea
G_{ab} \cV^a \tau^b  &= & \kappa_\cV \theta_{(\cV)} - \Lie_\cV \theta_{(\tau)}  + \tn \Lie_{\cV} C - d^2 C + 2  d_A ( C \tom^A) \label{Ttt}\\
& & - \left( \| \sigma_{(\ell)} \|^2 - C^2 \| \sigma_{(n)} \|^2 + \frac{1}{(n\!-\!1)} \left[\tl^2 - C^2 \tn^2 \right]  \right)  \nn \, ,
\eea
while from (\ref{dtom}) we obtain
\be
 e_A^a G_{ab} \tau^b  = \Lie_{\cV}  \tilde{\omega}_A  + \theta_{(\cV)} \tom_A - d_A \kappa_\cV + d^B k_{AB}^{(\tau)} -  d_A \tl - C d_A \tn 
\label{Tt} \, . 
\ee

Subject to these constraints, in future sections
we will view $( C, \tq_{AB}, \kappa_{\cV}, \tom_A, k^{(\ell)}_{AB}, k^{(n)}_{AB})$ as an (overdetermined) initial data set on $H$ (at least in the spacelike case) and then 
use them to perturbatively construct the nearby spacetime. 

\subsubsection{Deforming $H$ }
Our next step is to understand how the geometry of $H$ changes if it is deformed.  
As just seen, the intrinsic geometry of $H$ is specified by $(C, \tilde{\cV}^A, \tq^{AB})$ while the extrinsic geometry is also 
specified if we have knowledge of $(\kappa_{\cV}, \tom_A, k^{ (\ell)}_{AB}, k^{(n)}_{AB})$. Already from our earlier calculations we know how to 
find the deformations of $\tq_{AB}$, $k^{(\ell)}_{AB}$, $k^{(n)}_{AB}$ and $\tom_A$. Thus we just need to calculate the deformations of 
$C$, $\tilde{\cV}^A$ and $\kappa_{\cV}$. Once again we restrict out attention to deformations that are normal to the $S_v$ and so are of the form:
\be
X^a = \alpha \ell^a - \gamma n^a \, . 
\ee

For definiteness we will also need to fix the scaling gauge for the null vectors. 
We do this by tying that scaling to the foliation of $H$ and, once again, it is easiest to see how this works by considering 
a parameterization of $H$: $x^\alpha = \mathcal{Y}^\alpha (v, \theta^\A)$.  Then, as shown in FIG.~\ref{DefH}, $X^a$ 
infinitesimally deforms the original surface a coordinate distance $\Delta \rho$ via
\begin{equation}
\mathcal{Y}^\alpha (v, \theta^\A) \rightarrow \mathcal{Y}^\alpha (v, \theta^\A) + (\Delta \rho) X^\alpha (v, \theta^\A) \, , 
\end{equation}
transferring the coordinates along with the surface. In coordinate terms
\be
X = \frac{\partial}{\partial \rho} \, . 
\ee
The surface coordinates are Lie-dragged by $X$ so we have
\begin{equation}
[ X, e_\A] = 0 \; \; \mbox{and} \; \; [ X, \mathscr{V}] = 0 \, , \label{LieDrag}
\end{equation}
where, as before,
\be
\mathscr{V}^a = (\ell^a - C n^a) +  e^a_A \tilde{\cV}^A  \label{scrV}
 \, . 
\ee
The shift vector $\tilde{\mathcal{V}}^A$ is purely gauge (depending only on the choice of coordinate system on the surface) 
and so we usually simplify our calculations by choosing it to vanish on $H$ itself. However even if we do this, deformations will generally
cause it to become non-zero off of $H$.
From
\be
[ X, \mathscr{V}] = 0 \; \; \Longrightarrow \; \;  e^a_A \delta_X \tilde{\cV}^A = - \delta_X \cV^a \,  \label{deltaXtV}
\ee
it follows that $\ell_a \delta_X \cV^a = n_a \delta_X \cV^a = 0$ and so one can expand (\ref{deltaXtV}) with (\ref{XlXn}) to demonstrate that
\be
\delta_X \tilde{\cV}^A = d^A \gamma  + (C d^A \alpha - \alpha d^A C) - 2  ( \gamma - \alpha C) \tom^A \label{dXV} \, . 
\ee
This rate of change will usually be non-zero. 

The same set of calculations also give us  $\kappa_X$ and the deformation of $C$:
\be
\kappa_X =  \Lie_\cV \alpha + \alpha \kappa_{\cV}   \label{kappaX}
\ee
and
\be
\delta_X C =  \Lie_{\cV} \gamma + C \kappa_X - \gamma \kappa_{\cV}  \, .  \label{dXC}
\ee

We still need to find the variation of $\kappa_{\cV}$. This is most easily found by expanding the ``commutator'' 
$\delta_X \kappa_{\cV} - \delta_{\cV} \kappa_X$ using the regular tools. The result is:
\bea
\delta_X \kappa_{\cV} &=& \delta_{\cV} \kappa_X + (d^A C)(d_A \alpha)  \label{dXkappa} + 2 \tom^A ( -  d_A \gamma + \alpha d_A C - C d_A \alpha)    \\
& &+  ( \gamma - \alpha C) \left( 3 \tom^A \tom_A + \mathcal{R}_{abcd} \ell^a n^b n^c \ell^d \right)   \, . \nn
\eea
In terms of surface quantities and the Einstein tensor, the Riemann term can be rewritten as
\bea
\mathcal{R}_{abcd} \ell^a n^b n^c \ell^d =  \frac{\tilde{R}}{2} +  \tl \tn - k^{(\ell)}_{A B} k_{(n)}^{AB} - \frac{1}{2} \mathcal{R}_{ab} \tq^{ab} - \mathcal{R}_{ab} \ell^a n^b   \, . 
\eea

We now have formulae for calculating the deformation of all quantities defining the intrinsic and extrinsic geometry of $H$.

\subsection{Spacetime near $H$}
With these results in hand we can perturbatively construct spacetime close to $H$. We will do this using 
inward-oriented Gaussian null  coordinates and use the deformation results from the previous subsections to find the perturbed metric
components. As a concrete example, this construction is implemented for the Kerr-Newman spacetime in Appendix \ref{KerrEx}. 

We proceed as follows. Start with an $H$ that is foliated into $S_v$ by a coordinate system $\{ v, \theta^\A \}$ which has been chosen so that 
$\tilde{\mathcal{V}}^A = 0$. Scale the null normals to the $S_v$ so that $\cV = \ell - C n$. This fixes all of our gauge freedoms on $H$. The coordinate
system is then extended off of $H$ using the inward null geodesics that cross $H$ tangent to $n^a$. Assume an affine parameterization $\rho$ with the 
initial scaling set by
\be
n = \frac{\partial}{\partial \rho} \, . 
\ee
This $\rho$ will be our off-$H$ coordinate. Note that these coordinates, 
though clearly similar in spirit, are not identical with the standard Eddington-Finkelstein coordinates familiar from the Kerr-Newman
family of spacetimes. As is shown in Appendix \ref{KerrEx}, the geodesics in the standard system are not normal to the surfaces of 
constant $v$ on the horizon. 

Next, setting $\rho=0$ on $H$ we Lie-drag the coordinates $\{v, \theta^\A \}$ along the null geodesics to its other level 
surfaces. Thus our full $(n\!+\!1)$-dimensional set of coordinates is $\{v, \rho, \theta^A\}$. Because $\rho$ is an affine geodesic parameter it follows 
that our initial relations:
\be
n \cdot \mathscr{V}  = - 1 \; \; \mbox{and} \; \;  n \cdot e_\A = 0 
\ee
are conserved, though in general for $\rho \neq 0$ we lose the initial orthogonality between  $\mathscr{V}$ and the $e_\A$. 
The spacetime metric in this coordinate system
is then determined by a scalar function $C_\rho$, an $(n\!-\!1)$-dimensional vector function $\tilde{\cV}_\rho^\A$ and an 
$(n\!-\!1) \times (n\!-\!1)$-dimensional metric tensor function $\tq^\rho_{\A \B}$:
\be
ds^2  =  g_{\alpha \beta} dx^\alpha dx^\beta 
=
 -  2   dv d\rho  
+  (2C_{\rho} + \tq_{\A \B}^\rho \tilde{\cV}_\rho^\A \tilde{\cV}^B_\rho) dv^2   
 + 2  \tq^\rho_{\A \B} \tilde{\cV}_\rho^A dv d \theta^B +  \tq^\rho_{\A \B} d\theta^A d\theta^\B \nn \, . 
\ee
The subscript/superscript $\rho$s are included to differentiate these functions from those defined only as initial data on $H$: 
\be
C_0 = C \; , \; \; q^0_{\A \B} = q_{\A \B} \; \mbox{and} \; \; \tilde{\cV}_0^\A = \tilde{\cV}^\A = 0 \, .  
\ee
We Taylor-expand these metric-determining functions in $\rho$:
\bea
C_\rho &=& C + \rho C' + \frac{\rho^2}{2!} C'' + \frac{\rho^3}{3!}  C'''  \dots  \\
\tq^\rho_{\A \B} & = & \tq_{\A \B}  + \rho \tq'_{\A \B} + \frac{\rho^2}{2!}   \tq''_{\A \B} + \frac{\rho^3}{3!} \tq'''_{\A \B} \dots \nn \\
\tilde{\cV}_\rho^\A & = &  \phantom{\cV^A}  \rho \tilde{\cV}'^\A + \frac{\rho^2}{2!}  \tilde{\cV}''^\A + \frac{\rho^3}{3!}  \tilde{\cV}'''^\A  \dots \nn 
\eea
where primes indicate deformations in the $n$ direction (evaluated on $H$).  For example, 
\bea
\tq_{\A \B}'' = \left. \delta_n \delta_n \tq_{\A \B} \right|_H  \,. 
\eea
Since $\tilde{\cV}_0^\A = 0$ there is no zeroth order term in $\tilde{\cV}_\rho$. Then, to second order in $\rho$, the metric takes the form
\bea
ds^2 & = & \left\{ -  2 dv d \rho  + 2 C dv^2   + \tq_{\A \B} d \theta^\A d \theta^\B  \right\}  \label{metric_expansion} \\
& & + \rho \left\{ 2 C'  dv^2  + 2 \tq_{\A \B} \tilde{\cV}'^\B dv d \theta^\A  + \tq'_{\A \B} d \theta^\A d \theta^\B \right\} \nn \\
& & + \frac{\rho^2}{2} \left\{2 \left( C'' + \tq_{\A \B} \tilde{\cV}'^\A \tilde{\cV}'^\B \right) dv^2 + 2 \left(\tq_{\A \B} \tilde{\cV}''^\B + \tq'_{\A \B} \tilde{\cV}'^\B  \right) dv d \theta^\A + \tq''_{\A \B} d \theta^\A d \theta^\B \right\} \, .  \nn
\eea
We apply  our earlier results to calculate each term of the expansions.  

With $X = n$ we have $\alpha = 0$ and $\gamma = -1$ and so we are dealing with a particularly simple deformation. 
First, by (\ref{kappaX}) this means that $\kappa_n = 0$. Then (\ref{dXC}), (\ref{dXq}) and (\ref{dXV}) respectively imply that:
\bea
C' &=&    \kappa_\cV  \, , \\
\tq'_{\A \B} &=&   2 k^{(n)}_{\A\B} \; \;  \mbox{and} \\
\tilde{\cV}'^A &=&    2 \tom^\A  \, . 
\eea
Next (\ref{dXkappa}), (\ref{dkn}) and (\ref{dtom}) can be used to obtain the  second order derivatives:
\bea
C'' &=& - 3 \tom^\A \tom_\A  -  \frac{\tilde{R}}{2} - \tl \tn + k^{(\ell)}_{\A\B} k_{(n)}^{\A\B} + \frac{1}{2} \mathcal{R}_{\alpha \beta} \tq^{\alpha \beta} +  \mathcal{R}_{\alpha \beta} \ell^\alpha n^\beta \, ,  \label{d2XC} \\
\tq_{\A \B}'' &  = & 2  k^{(n)}_{\A\C} k^{(n) \C}_\B 
  - 2 e_\A^\alpha n^\beta e_\B^\gamma n^\delta C_{\alpha \beta \gamma \delta} 
  - \frac{2}{(n\!-\!1)} \tq_{\A\B} \mathcal{R}_{\gamma \delta} n^\gamma n^\delta  \label{d2Xq}  \; \; \mbox{and} \\
\tilde{\cV}''^\A & = &  2 d_\B k_{(n)}^{\A \B} -2  d^\A \tn - 2 \tn \tom^\A -  4 k_{(n)}^{\A \B} \tom_\B - 2 e^{\A \alpha} \mathcal{R}_{\alpha \beta} n^\beta \, .
\label{d2XV}
\eea

Often it will be most convenient to leave the expanded metric in the form (\ref{metric_expansion}). However we can also combine our results to 
write it explicitly as
\bea
ds^2 & \approx & \left\{ - 2 dv d \rho +  2C dv^2  +  \tq_{\A \B} d\theta^A d\theta^\B \right\} 
 +2  \rho \left\{  \kappa_\cV dv^2 + 2 \tom_\A dv d \theta^\A +  k^{(n)}_{\A \B} d\theta^\A d \theta^\B \right\}  \label{metric_expansionII} \\
& & + \rho^2 \left\{
\begin{array}{l}  \left( -  \frac{\tilde{R}}{2}  +  \tom^\A \tom_\A - \tl \tn + k^{(\ell)}_{\A \B} k_{(n)}^{\A \B} + \frac{1}{2} \mathcal{R}_{\alpha \beta} \tq^{\alpha \beta} + \mathcal{R}_{\alpha \beta} \ell^\alpha n^\beta \right) dv^2  \\
\\
+2 \left( d_\B k_\A^{(n) \B} - d_\A \tn   - \tn  \tom_\A 
- e_\A^\alpha \mathcal{R}_{\alpha \beta} n^\beta \right) dv d\theta^\A \\ \\
+ \left(k^{(n)}_{\A\C} k^{(n) \C}_\B 
  - e_\A^\alpha n^\beta e_\B^\gamma n^\delta C_{\alpha \beta \gamma \delta} 
  - \frac{1}{(n\!-\!1)} \tq_{\A\B} \mathcal{R}_{\gamma \delta} n^\gamma n^\delta \right) d\theta^\A d\theta^B
\end{array}
 \right\} \nn \, . 
\eea
This is the metric that we will use in future sections where we construct spacetimes close to horizons. In doing that we, of course, assume
that these first few terms of the asymptotic series provide a good approximation to the true metric (for small $\rho$). However we will not
rigourously address issues of convergence for the full series. 

Note too the appearance of the Weyl term at second order. In general this is new data and not specified by the set:
$( C, \tq_{AB}, \kappa_{\cV}, \tom_A, k^{(\ell)}_{AB}, k^{(n)}_{AB})$. In our examples, it can be neglected but we will 
return to it in our final Discussion. 

\section{Horizons}
\label{horizons}

Before constructing near-horizon spacetimes, we recall some horizon definitions and properties. There are two principal types:
geometric and causal. Geometric horizons 
are quasilocally defined and include trapping horizons, isolated horizons and dynamical horizons. Like apparent horizons these are defined in terms 
of the geometry of $(n\!-\!1)$-surfaces and are closely related to trapped surfaces (for which both $\tn < 0$ and $\tl < 0$). In contrast, event horizons
are defined relative to the causal structure of the full spacetime. This makes them in one way simpler than geometric horizons: one can identify them 
with just a knowledge of how to calculate null geodesics. However in another way they are significantly more complicated: they are highly non-local 
and defined by the future behaviour of null geodesics. For more details and a discussion of the 
properties of these objects see review articles such as \cite{LivRev, isoreview, gourgoul, tomasIso, BHboundaries, SEHreview} or one of the  
original sources as cited below. All of these results are also discussed in some detail in \cite{bfbig} in the same style that we use in this paper. 

\subsection{Geometric horizons}

\subsubsection{General cases}
In an $(n\!+\!1)$-dimensional spacetime $(M,g_{ab}, \nabla_a)$,  a \emph{future outer trapping horizon (or FOTH)} is a $n$-surface $H$ that is 
foliated by  spacelike $(n\!-\!1)$-surfaces $(S_v, \tq_{AB}, d_A)$ such that on each surface: i) $\tl = 0$, ii) $\tn < 0$ and iii) there is a positive function $\beta$ such that
$\delta_{\beta n} \tl < 0$ \cite{hayward}. 
These conditions are intended to (locally) mimic those used to define apparent horizons \cite{hawkellis}: each slice of a FOTH is marginally 
outer trapped ($\tl = 0$) and the other two conditions guarantee that it is possible to deform the $S_v$ inwards so that they become fully trapped. 


As in our construction of section \ref{GenSurf}, we can define an evolution vector field 
\be
\mathcal{V}^a = \ell^a - C n^a \, , 
\ee
that is both normal to the leaves of the foliation and maps them into each other.
%
Then with $\tl = 0$ for each $S_v$, it follows that $\Lie_\cV \tl = 0$. Using (\ref{explderiv}) this may be expanded into a second order differential 
equation for $C$ over each $S_v$. The assumptions that the dominant energy condition holds on the horizon
and that $\delta_{\beta n} \tl < 0$ (for some $\beta$) may be used in a maximum principle argument to show that $C \geq 0$. 
In turn this means 
\be
\Lie_\cV \vS = - C \tn \vS \geq 0 \, , \label{SecondLaw}
\ee
since $\tn < 0$. Under these conditions the FOTH is spacelike or null and the area is non-decreasing. 
This is the second law of FOTH mechanics \cite{hayward}. 

If $C=0$ then the FOTH is null and a type of \emph{isolated horizon} (specifically a non-expanding horizon) \cite{isoMech,isoGeom}. 
Some of the properties of isolated horizons will be 
discussed in section \ref{IsoPhase}. Here we simply note that the 
intrinsic and extrinsic geometries of isolated horizons are unchanging in time and there is no flux of stress-energy or gravitational waves 
across their horizons: they are equilibrium states. By the zeroth law of isolated horizon mechanics $\kappa_{\ell}$ is constant over an isolated
horizon. All Killing horizons are examples of isolated horizons.

If $C>0$ then a FOTH is a  \emph{dynamical} horizon\cite{ak} and, as noted, is spacelike and expands in area. As for isolated horizons, we only 
summarize properties that are relevant to the current discussion. By another maximum principle argument, it can be demonstrated
that the foliations of a dynamical horizon are unique: there is only one foliation for which $\tl = 0$ \cite{Ashtekar:2005ez}. This is
very convenient for our discussions as we do not need to worry about whether or not geometric properties are foliation dependent. They 
are, but since the foliation is unique this is fine! On a closely related note it can be shown that if a FOTH 
transitions from being isolated to dynamical it does so all at once. That is, on each $S_v$, $C$ is either zero everywhere or it is zero nowhere
\cite{Ashtekar:2005ez}.

Dynamical FOTHs (like dynamical apparent horizons)
\emph{are not} uniquely defined. Though the $S_v$ cannot be deformed within $H$, they can be deformed out of $H$ (see \cite{Ashtekar:2005ez, bfbig} for theoretical discussions or \cite{Nielsen:2010wq} for concrete demonstration within the Vaidya spacetime). At a local level this 
non-uniqueness manifests itself in the fact that, for a given $S_v$, changing the scaling of the null vectors will cause equation (\ref{explderiv})  
to solve for a different $\mathcal{V}^a$ which in turn will evolve that $(n\!-\!1)$-surface into an $H' \neq H$. By contrast isolated FOTHs 
\emph{are} rigid: for that case it can be shown that the only allowed deformations are those that change the foliation of $H$ but do not 
otherwise affect it \cite{Ashtekar:2005ez, bfbig}. 

For the rest of this paper we will consider FOTHs that satisfy the dominant energy condition and so are null or spacelike. 
We respectively refer to them as isolated
or dynamical FOTHs. There are also more exotic forms of non-FOTH trapping horizons associated with apparent horizon 
``jumps'' (see for example \cite{bendov, mttpaper,chu}), inner horizons or white holes \cite{hayward}. However, they can be left aside for the purposes
of this article. 

\subsubsection{Slowly evolving horizons}
\label{SEH}

With isolated horizons characterizing equilibrium states we can turn our attention to the near-equilibrium regime. Intuitively these should be 
nearly isolated and so nearly null with quantities on the horizon changing slowly in time. However, given that dynamical horizons are naturally spacelike,
there is no real notion of time on the horizon and so the trick is invariantly characterizing this intuition. We do this in the definition below by
defining a ``slowness'' parameter $\varepsilon$ and then using it as a basis of comparison for tracking rates of evolution up $H$. 
The characterization of a \emph{slowly evolving horizon} given below is simplified but also slightly strengthened from that originally 
developed in \cite{prl, bfbig} and recently reviewed in \cite{SEHreview}. More motivation for the definition can be found in those references. 

\vspace{0.3cm}
\noindent
{\bf Definition:} Let $\triangle H = \{ \cup_v S_v : v_1 \leq v \leq v_2\}$ be a section of a FOTH with evolution vector field 
$\cV^a = \ell^a - C n^a$. Define an \emph{evolution parameter} $\varepsilon$ via
\be
\varepsilon^2/R_H^2 =  \mbox{Maximum} \left[ C \left( \norm \sigma^{(n)} \norm ^2 + \mathcal{R}_{ab} n^a n^b + \tn^2/2 \right) \right]
\label{SEHCond}
\ee
where $R_H$ is the characteristic length scale for the problem. If $\varepsilon \ll 1$ and the foliation parameter has been chosen so that 
$\left\| \cV \right\| = \sqrt{2 C} \lesssim \varepsilon$, then we say that $\triangle H$ is a \emph{slowly evolving horizon (SEH)} if on each 
$S_v$: 
\begin{enumerate}[(a)]
\item the dominant energy condition holds, \label{a}
\item $\displaystyle |\tilde{R}|, \tom_A \tom^A, | d_A\tom^A |$ and $\mathcal{R}_{ab} \tq^{ab} \lesssim 1/
R_H^2$, \label{b}
\item  derivatives of a horizon field tangent to the $S_v$ are at most of the same magnitude as the maximum of the
 original field. For example, $\displaystyle \norm d_A \tn \norm \lesssim \tn^{max}/R_H$ and \label{c}
\item  derivatives of a horizon field ``up'' the horizon in the $\mathcal{V}^a$ direction are an order of magnitude $\epsilon^2/R_H$ 
smaller than the maximum of the original field. For example $\displaystyle | \Lie_\cV \kappa_\cV |  \lesssim (\varepsilon^2/R_H) \kappa^{max}_{\cV}$
and $|\Lie_{\cV} C| \lesssim (\varepsilon^2/R_H) C^{max} $. \label{d}
\end{enumerate}
The notation $X \lesssim Y$ means that $X \leq k_o Y$ for some constant $k_o$ of order one while the superscript 
$max$ indicates the largest absolute value quantity over  $S_v$ for the quantity to which it is attached. \\

Let us consider the definition in a bit more detail. First, for standard black holes the 
characteristic scale is the areal radius of the horizon while for black brane spacetimes in an AdS background, it is the radius of curvature 
of that background. Next, note that the definition of the evolution parameter
is scaling independent.  Geometrically it has the implication that the rate of change of the area (or volume depending on the dimension) 
element is small relative to changes in proper length measured up the horizon:
\be
\Lie_{\widehat{\cV}} \vS \leq -  \sqrt{\frac{C}{2}} \tn \vS \, . 
\ee 
If this invariant condition is met, then by equation (\ref{ReLabel}) the foliation labelling may always be chosen so that $\| \cV \| \lesssim \varepsilon$
and so that requirement is not independent of condition (\ref{SEHCond}). Among
other things, this choice ensures that in any approach to isolation, $C$ approaches zero in the expected way.  
Turning to the remaining clauses the energy condition (\ref{a}) has the usual physical implications for the matter fields while 
(\ref{b}) and (\ref{c}) restrict the geometry on the surface to be not too extreme: extreme conditions will generally mean that the horizon will not remain 
slowly evolving for long. 

Finally (\ref{d}) demands that geometric properties of the horizon change slowly relative to $\mathcal{V}$:
this is the clause modified from earlier definitions. For simplicity we have opted for a general statement of principal here rather than 
a list of specific quantities that must meet this definition. We have also strengthened the statement by requiring that ``time-derivatives'' of 
quantities be order $\varepsilon^2$ smaller than the original quantities rather than order $\varepsilon$. This stronger statement continues
to be consistent with known examples of slowly evolving horizons.
It can also be reasonably argued that $\varepsilon^2$ (rather than $\varepsilon$) is the true scale of slowness for the problem rather than 
$\varepsilon$. 
It is $\varepsilon^2$ that appears in (\ref{SEHCond}) and the rate of change of the area/volume element is $\varepsilon^2$. This strengthened statement is required for our upcoming demonstration that there is an event horizon candidate close to 
any SEH. 

An important physical property of SEHs, which supports their interpretation as near-equilibrium states, is that they obey versions of the zeroth and first 
laws of black hole mechanics\cite{prl,bfbig}. For a near-equilibrium state, one would expect the surface gravity $\kappa_{\cV}$ to be approximately 
constant  and indeed this follows from equation (\ref{dtom}). Variations are at order $\varepsilon^2$:
\be
\| d_A \kappa_\cV \| \sim \frac{O(\varepsilon^2)}{R_H^2} \label{zeroth}
\ee
(thanks to the strengthening of (\ref{d}) this is also slightly stronger than the equivalent result in \cite{prl, bfbig}). 
One can also combine (\ref{explderiv}) and (\ref{expnderiv}) 
to derive a first law:
\begin{equation}\label{FirstLaw}
  \frac{ \kappa_o \dot{a}}{8 \pi G} \approx 
\int_{S_v} \vS \left\{\frac{\norm\sigma^{(\ell)}\norm^2}{8 \pi G} 
  + T_{ab} \ell^a \ell^b \right\}  \, , 
\end{equation}
where we have applied the Einstein equations to turn Ricci terms into stress-energy ones\footnote{Though this is generally referred to as 
a dynamical first law of black hole mechanics it has been pointed out (see for example \cite{GourJara}) 
that it is in closer analogy with the Clausius form of the second law of thermodynamics. 
In particular, unlike either the usual first law of thermodynamics or the first law of black hole mechanics for stationary black holes, this law makes no reference to internal energy. }.
%

\subsection{Causal horizons}


\subsubsection{General case}

The alternative to geometric horizons are causally defined event horizons. 
It is well known that these are teleological -- their position depends on future
events. This follows directly from their definition and is probably most easily understood with the help of FIG.~\ref{EHFig}.
\begin{figure}
\scalebox {1.1}{\includegraphics{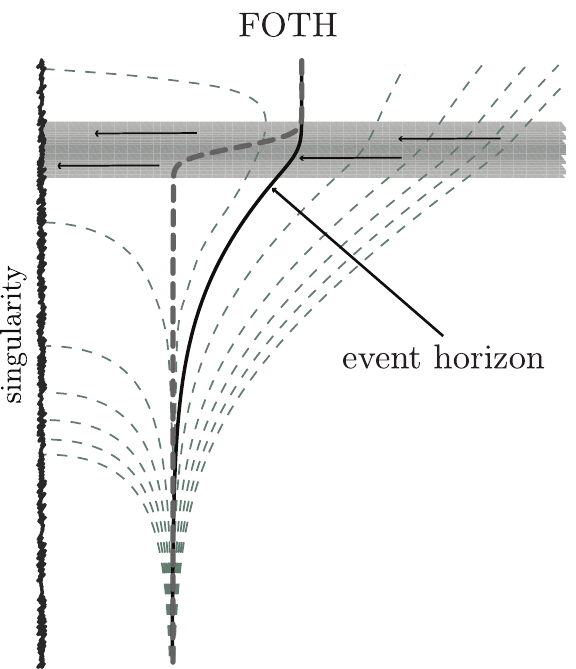}}\caption{A schematic that plots both the (spherically symmetric) FOTH  and event horizon for a 
typical Vaidya spacetime in which a shell of dust (the shaded gray region) falls into a pre-existing black hole. 
In this figure, horizontal position records the areal radius of 
the associated spherical shell while the direction of increasing time is roughly vertical outside the event horizon but 
tipping horizontal-and-to-the-left inside. On both sides, inward-moving null geodesics are horizontal while ``outward-pointing'' null geodesics 
are represented by gray dashed lines.}
\label{EHFig}
\end{figure}
An event horizon is the boundary of a \emph{causal black hole}: a region of spacetime from which no causal signal can escape. 
Such a surface is necessarily null and, for outside observers, it is the boundary between the unobservable events inside the black hole and 
those outside that can be seen. One determines the extent (or existence) of a black hole by tracing all causal paths ``until the end of time'' and then
retroactively identifying any black hole region. The exterior of the black hole is the set of all points for which at least one causal signal reaches 
$\mathscr{I}^+$ (future null infinity) while the interior is the set of all points from which no signal escapes. The boundary between the two regions 
is the event horizon and it is necessarily a congruence of null geodesics. 
 
In practice, of course, one cannot trace the paths of all possible null geodesics (let alone all causal curves). However, in the case where a spacetime
ultimately settles down to an equilibrium state, there is a short-cut to finding the event horizon \cite{thornburg}. By the uniqueness theorems the only 
$(3\!+\!1)$ dimensional, asymptotically flat, stationary and vacuum black hole spacetimes are members of the Kerr family. For these black holes,
 the location of the event horizon is well-known
and so once that equilibrium state is reached, one can trace its evolution back into the past to find its location at all times (FIG.~\ref{EHFig} again). 

Given the nature of their definition, it is perhaps no surprise that event horizons have some unusual properties. A particularly important one
is that infalling matter can curtail rather than drive the expansion of an event horizon (also demonstrated in FIG.~\ref{EHFig}). 
If one thinks of an event horizon as a standard object this behaviour seems counter-intuitive, however 
it is straightforward to see how this happens. 

The event horizon is a null surface and so is ruled by null geodesics. 
In this case (\ref{explderiv}) reduces to the Raychaudhuri equation and can be rewritten as
\begin{align}
\kappa_{\ell} \tl - \Lie_{\ell} \tl =  \frac{1}{2} \tl^2  + \norm \sigma_{(\ell)}\norm^2 + \mathcal{R}_{ab} \ell^a \ell^b   \label{explderiv2} \, .
\end{align}
Assuming the Einstein equations and null energy condition the right-hand side of this equation is non-negative. The characteristic evolution of 
the horizon is then determined by the relative magnitudes of $\kappa_\ell \tl$ and $\Lie_\ell \tl$. If $\Lie_\ell \tl$ dominates, then $\tl$ necessarily 
decreases with time and, in particular, infalling matter further curtails the rate of expansion. Though at first thought this might seem strange, on second
thought it makes sense: the equation is simply telling us that the gravitational influence of more mass inside the horizon decreases the rate
of expansion of the horizon, just as it would for any other set of outward moving geodesics. 

However, in a regime where $\kappa_\ell \tl$ dominates then we have a more naively intuitive situation. In that case an increase in the right-hand
side of (\ref{explderiv}) will drive a corresponding increase in $\kappa_\ell \tl$. In particular, if $\kappa_\ell$ is constant or non-increasing, then it is 
the rate of expansion $\tl$ that must increase.  
This behaviour is seen in  perturbative calculations \cite{PertEx} and we now demonstrate that it is to be expected for near-equilibrium event horizons.

\subsubsection{Slowly evolving null surfaces}

\noindent
{\bf Definition:} Let $\triangle H$ be a section of an $n$-dimensional null surface with tangent vector field $\ell^a$ with 
characteristic scale $R_H$. Then we say that $\triangle H$ is a \emph{slowly evolving null surface (SENS)} if for some small $\varepsilon^2$:
\begin{enumerate}[(a)]
\item $\frac{1}{2} \tl ^2 \lesssim \varepsilon^2 \left( \norm \sigma_{\ell} \norm^2 + \mathcal{R}_{ab} \ell^a \ell^b \right)$ and \label{a2}
\item $\ell^a$ can be scaled so that $\kappa_\ell$ is of order $1/R_H$ and
\be
\Lie_\ell \tl \lesssim \left( \frac{\varepsilon^2}{R_H} \right) \tl \, .
\ee  \label{b2}
\end{enumerate}
(In order to ensure future compatibility with our discussion of SEHs we have written our expansion parameter as $\varepsilon^2$ 
rather than $\varepsilon$). \\

The application of these conditions is straightforward. First 
it is straightforward to see that scaling invariant condition (\ref{a2}) means that the $\tl^2$ term in the right-hand side of (\ref{explderiv2}) can be
neglected while by (\ref{b2}) we can scale the null vectors 
so that on the left-hand the $\Lie_\ell \tl$ term may be also dropped. Then we have 
\be
\kappa_\ell \tl \approx  \norm \sigma_{(\ell)} \norm^2 + \mathcal{R}_{ab} \ell^a \ell^b 
\ee
and if $\kappa_\ell > 0$, the expansion is positive and driven by the flux terms on the right-hand side. 

These are exactly the kind of conditions that one might expect to hold in a perturbative near-equilibrium regime. If the rate of expansion is 
small then its square will be even smaller while in perturbative calculations one often assumes a derivative expansion where time 
derivatives of quantities are always much smaller than the quantities themselves. This is the same kind of assumption as we made in our 
definition of slowly evolving horizons.

\section{Extremal isolated horizons and near horizon spacetimes}
\label{Extremal} 

With the mathematical formalism set up and definitions established we are now ready to consider applications. 
The first will be a study of spacetime around an extremal isolated horizon. We demonstrate that at leading order
the metric near an extremal horizon is a near-horizon spacetime of the type studied in the near-horizon
literature (for example \cite{hariOld, hariClass,hariNewer}). As a first step to that end, we consider the phase
space of isolated horizons.

\subsection{Phase space of isolated horizons}
\label{IsoPhase}
Following \cite{isoGeom}, an isolated horizon is an $n$-dimensional null surface with null tangent vector $\ell^i$ and induced 
(null) metric $\tq_{ij}$. Its intrinsic geometry is time invariant and so 
\be
\Lie_\ell \tq_{AB} = 0  \; \; \Longrightarrow \; \; k^{(\ell)}_{AB} = 0 \; \; \Longrightarrow \; \; \tl = 0 \; \mbox{and}  \; \sigma^{(\ell)}_{AB} = 0 \, . 
\label{isoCond1}
\ee
However its extrinsic geometry is also invariant which give us further constraints on possible on-horizon data. 
From 
$
\Lie_\ell k^{(\ell)}_{AB} = 0
$
it follows from (\ref{dkl}) and (\ref{explderiv}) that 
\be
\mathcal{R}_{a b} \ell^a \ell^b = 0 \; \; \mbox{and} \; \; e_A^a \ell^b e_c^C \ell^d C_{abcd} = 0  \label{Lklz} \, . 
\ee 
Alternatively the restriction on the Ricci tensor follows from (\ref{Ttt}) and (\ref{TVt}) -- the Hamiltonian constraint plus part of the momentum constraint. The rest of the momentum constraint (\ref{Tt}) reduces to the zeroth law
\be
d_A \kappa_\ell = 0 \label{dkappa}
\ee 
and a further constraint on the $(n\!+\!1)$-Ricci tensor
\be
e_A^a R_{ab} \ell^b = 0 \, .  \label{isoCond4}
\ee
Finally in most cases, $k^{(n)}_{AB}$ is fully determined by other quantities.  From 
$\Lie_\ell k^{(n)}_{AB} = 0$ and (\ref{dkn}) we find that
\bea
 \kappa_\ell k^{(n)}_{ AB} +   \frac{1}{2} \tilde{R}_{AB}   =  
  \frac{1}{2} e_A^a e_B^b \mathcal{R}_{ab} 
  + d_{(A} \tom_{B)} + \tom_A \tom_B  \label{IH_constraint} \, , 
\eea
which for $\kappa_\ell \neq 0$ can be solved as:
\bea
 k^{(n)}_{ AB}    =  \frac{1}{\kappa_\ell}
 \left(  \frac{1}{2} e_A^a e_B^b \mathcal{R}_{ab} 
  + d_{(A} \tom_{B)} + \tom_A \tom_B -    \frac{1}{2} \tilde{R}_{AB} \right)  \, . 
\eea
These are our constraints on the phase space of possible isolated horizons -- that is 
the allowed values of $\{\tq_{AB}, k^{(n)}_{AB}, \tom_A, \kappa_\ell, R_{ab}, C_{abcd} \}$ on $H$. 

The fit between isolated horizon constraints and the Hamiltonian and momentum constraints for spacelike surfaces is not perfect, but this is 
not surprising as we are now dealing with a somewhat specialized null surface rather than a general spacelike surface. In particular,
data on a single null surface will never be sufficient to fully specify even a neighbourhood of that surface:
the domain of dependence of a null surface is always empty. Perhaps the best demonstration of this fact is that there exist
spherically symmetric isolated horizons that have the same geometry as Schwarzschild horizons but live in spacetimes which, even arbitrarily close to the horizon, are globally different \cite{lewandowski,Ashtekar:2000sz}.  
To get a proper initial value formulation one must specify data on 
two $n$-dimensional null surfaces that intersect along an $(n\!-\!1)$-dimensional surface \cite{badri, lewandowski, friedrich}. For our example
one would need: $k^{(\ell)}_{AB}$, $k^{(n)}_{AB}$, $\tom_A$ on an $S_v$ along with $e_A^a \ell^b e_C^c \ell^d C_{abcd}$ on $H$ and
$e_A^a n^b e_C^c n^d C_{abcd}$ on an inward-moving null surface that intersects $H$ along $S_v$ (for example a congruence of inward
moving null geodesics). 

So, in general the data specified on the isolated horizon do not fully specify the behaviour of spacetime, even in a restricted 
neighbourhood. That said if we assume that our expansion around the horizon is a good approximation to the full spacetime 
(with leading order terms dominating those at subleading order)
and work at small $\rho$ so that higher order terms can be neglected, then we do not need exact knowledge of the Weyl components.  
They are only required to evaluate higher-order terms in the series.

\subsection{Extremal horizons and near-horizon spacetimes}
There are several equivalent ways to characterize extremal isolated horizons \cite{ExtremalPaper} but for our purposes it is most convenient to define
them as the subset of isolated horizons for which the surface gravity vanishes: $\kappa_\ell = 0$. Then (\ref{dkappa}) is trivially satisfied while 
(\ref{IH_constraint}) reduces to 
\bea
  \frac{1}{2} \tilde{R}_{AB}   =  
  \frac{1}{2} e_A^a e_B^b \mathcal{R}_{ab} 
  + d_{(A} \tom_{B)} + \tom_A \tom_B  \label{Ex_constraint}
\, \, , 
\eea
so that the extrinsic curvature $k^{(n)}_{AB}$ decouples from the constraints and becomes freely specifiable data. 
%

Applying $\tl = \kappa_\ell = 0$ and $\sigma^{(\ell)}_{AB} = 0$ to (\ref{metric_expansionII}) the spacetime near a extremal isolated horizon takes the form:
\bea
ds^2 & \approx & \phantom{+2 \rho} \Big\{ - 2 dv d \rho  +  \tq_{\A \B} d\theta^A d\theta^\B \Big\} \label{IsoExpansion}  \\
&&  +2  \rho \left\{   2 \tom_\A dv d \theta^\A +  k^{(n)}_{\A \B} d\theta^\A d \theta^\B \right\}  \nn \\
& & + \rho^2 \left\{
\begin{array}{l}  \left(  2\tom_\A \tom^\A  + {d}_{\A} \tilde{\omega}^{\A} - \mathcal{R}_{\alpha \beta} \ell^a n^b \right) dv^2  \\
\\
+2 \left(2 k_{\A \B}^{(n)} \tom^\B + d_\B k_\A^{(n) \B} - d_\A \tn - \tom_\A \tn - e_\A^\alpha \mathcal{R}_{\alpha \beta} n^\beta \right) dv d\theta^\A \\ \\
+ \left(k^{(n)}_{\A\C} k^{(n) \C}_\B 
  - e_\A^\alpha n^\beta e_\B^\gamma n^\delta C_{\alpha \beta \gamma \delta} 
  - \frac{1}{(n\!-\!1)} \tq_{\A\B} \mathcal{R}_{\gamma \delta} n^\gamma n^\delta \right) d\theta^\A d\theta^B
\end{array}
 \right\} \nn \, ,
\eea
where the trace of (\ref{Ex_constraint}) has been used to simplify the $dv^2$ term that is proportional to $\rho^2$.
If we retain only the leading order terms in each coefficient this becomes:
\bea
ds^2 & \approx &
 \rho^2
  \left(  2\tom_\A \tom^\A  + {d}_{\A} \tilde{\omega}^{\A} - \mathcal{R}_{\alpha \beta} \ell^a n^b \right) dv^2  
- 2 dv d\rho + 4 \rho \tom_\A dv d \theta^\A  +  \tq_{\A \B} d\theta^A d\theta^\B  \, . \label{leadingOrder}
\eea

For many readers, this may be a familiar expression. There is a large literature (see \cite{hariOld, hariClass,hariNewer} and references therein) on extremal black holes and their 
near-horizon properties; this metric appears frequently in that work. Very briefly it arises in the following way. 
Start with an extremal black hole in a stationary $(n\!+\!1)$-dimensional spacetime. In a neighbourhood of the horizon  \cite{hariOld} 
that the spacetime metric can be written in the form \cite{hariOld}:
\be
ds^2 = r^2 F(r, \theta^\B) d\tilde{v}^2 + 2 d\tilde{v} dr + 2 r h_\B (r,\theta^\A) d\tilde{v} d \theta^B + \tq_{\B \C} (r, \theta^\A) d\theta^\B d\theta^\C \, .
\label{baseMetric}
\ee
Because the metric is stationary there is no $v$ dependence in any of the terms and the ``radial'' coordinate $r$ measures (affine)
coordinate distance from the horizon along null geodesics. The $r^2$ in front of the $dv^2$ identifies the horizon as extremal. 
Making the coordinate transformation 
\be
r = \epsilon \rho  \; \; \mbox{and} \; \;  \tilde{v} = \frac{v}{\epsilon}  \, , 
\ee
the near-horizon limit consists of sending $\epsilon \rightarrow 0$. Then (\ref{baseMetric}) becomes
\be
ds^2 = \rho^2 F(0,\theta^\A) dv^2 + 2 dv d\rho + 2 \rho h_\B(0,\theta^\A) dv d\theta^\B + \tq_{\B \C}(0,\theta^\A) d\theta^\B d\theta^\C \, . \label{nearHorizon}
\ee
Viewing this limit as a spacetime in its own right, the Einstein equations reduce to 
\be
\tilde{R}_{\A \B} = \frac{1}{2} h_\A h_\B - d_{(\A} h_{\B)} + \Lambda \tq_{\A \B} \label{NHconstraint}
\ee
and 
\be
F = \frac{1}{2} h_\A h^\A - \frac{1}{2} \nabla_\A h^\A + \Lambda \, . 
\ee
The equivalence of (\ref{baseMetric}) with our leading-order spacetime (\ref{leadingOrder}) is obvious. For 
\be
r = - \rho \; \; \mbox{and} \; \; h_\A =-  2 \tom_\A 
\ee
the forms of the metrics match exactly, as do their defining constraint equations (\ref{Ex_constraint}) and (\ref{NHconstraint}).
 
To leading order the spacetime near a general extremal isolated horizon (including one that might be embedded in a non-stationary spacetime) is 
the same as that near an extremal black hole which is embedded in a stationary spacetime. 

 
\section{Event horizon candidates near slowly evolving horizons}
\label{EventHorizon}

Next we show that there is always a slowly evolving null surface in close proximity to any slowly evolving horizon. In the case where the geometric
horizon remains slowly evolving for the rest of time, this null surface is the event horizon. To demonstrate this we will use our earlier results to 
construct the 
spacetime close to SEH and then search for an event horizon candidate in that  neighbourhood. 


\subsection{Locating the event horizon candidate}

As for isolated horizons, we base our construction on inward-moving null geodesics with the $(v, \rho, \theta^\A)$ coordinate system and 
work with the metric (\ref{metric_expansion}).
$C$ sets our scale-of-smallness and we consider surfaces
$\mathcal{E}$ that can be defined by a series of the form:
\be
\rho^{(\mathcal{E})} \approx \rho_{(1)}(v, \theta^\A) + \rho_{(2)} (v, \theta^\A) + \rho_{(3)} (v, \theta^\A) + \rho_{(4)}(v, \theta^\A) \dots \label{HorLoc}
\ee
where $\rho_{(J)}  \sim C^{J}$ and we assume that (similar to other SEH quantities and subject to the inclusion of appropriate powers of 
scaling factor $R_H$)
\be
\frac{d \rho_{(J)}}{dv} \lesssim C^{J+1} \; \; \mbox{and} \; \;   \| d_A \rho_{(J)} \| \lesssim C^J  \, . 
\ee
Then, to second order, the induced metric on $\mathcal{E}$ is
\bea
d\Sigma^2 &=& \tq^H_{\A \B} d \theta^\A d \theta^\B  \label{EH_metric}\\
&& + \left\{ 2 \left( C +   \rho_{(1)} C' \right) dv^2 + 2 \left(\rho_{(1)} \tq_{\A \B} \tilde{\cV}'^\B - d_\A \rho_{(1)} \right) dv d \theta^\A + \left(\rho_{(1)} \tq'_{\A \B} \right) d\theta^\A d \theta^\B \right\} \nn \\
&& + \Biggr\{ \biggr(2 \rho_{(2)} C' - 2 \dot{\rho}_{(1)} +  \rho_{(1)}^2 C'' + \rho_{(1)}^2 \tq_{\A \B} \tilde{\cV}'^\A \tilde{\cV}'^\B \biggr) dv^2  \nn \\
&& \qquad
 +  2 \left( - d_\A \rho_{(2)} + \rho_{(2)} \tq_{\A \B} \tilde{\cV}'^\B + \frac{1}{2} \rho_{(1)}^2 \tq_{\A \B}' \tilde{\cV}'^\B + \frac{1}{2} \rho_{(1)}^2 \tq_{\A \B} \tilde{\cV}''^\B \right) dv d \theta^\A  \nn \\
& & \qquad \qquad + \left(\rho_{(2)} \tq'_{\A \B} + \frac{1}{2} \rho_{(1)}^2 \tq_{\A \B}'' \right) d\theta^\A d \theta^\B \Biggr\} \nn \, 
\eea
where  $ \dot{\rho}_{(1)} = \Lie_\cV \rho_{(1)} = {d \rho_{(1)}}/{dv}$. 

Any event horizon candidate will be null and so we need to solve for the $\rho_{(J)}$ so that the determinant of this induced metric vanishes. 
To that end recall that a general metric of the form
\bea
d\Sigma^2 = F dv^2 + 2 V_\A dv d \theta^\A + h_{\A \B} d \theta^\A d \theta^\B
\eea
has determinant
\be
(F - h_{\A \B} V^\A V^\B ) \times \det (h) \, . 
\ee
If $h_{\A \B}$ is spacelike (as $\tq_{\A \B}$ is in our case) then the full metric determinant vanishes if and only if 
\be
F - h_{\A \B} V^\A V^\B = 0 \, . 
\ee
We can apply this to the induced metric (\ref{EH_metric}) and solve order-by-order for the $\rho_{(J)}$. To zeroth order any such $\mathcal{E}$ is 
already null but at first order we must have
\be
C + \rho_{(1)} C' = 0 \, \,  
\ee
and so find that 
\be
\rho_{(1)} = - \frac{C}{\kappa_\cV} \, . 
\label{rho1}
\ee
The determinant also vanishes to second order if
\bea
 \biggr(2 \rho_{(2)} C' - 2 \dot{\rho}_{(1)} +  \rho_{(1)}^2 C'' + \rho_{(1)}^2 \tq_{\A \B} \tilde{\cV}'^\A \tilde{\cV}'^\B\biggr) 
   - \tq^{\A \B} \left( \rho_{(1)} \tq_{\A \C} \tilde{\cV}'^\C - d_\A \rho_{(1)} \right) \left(\rho_{(1)} \tq_{\B \D} \tilde{\cV}'^\D - d_\B \rho_{(1)}  \right)  = 0 \,  . 
\eea
This is easily solved for $\rho_{(2)}$ 
\bea
\rho_{(2)}  =  \frac{1}{C'} \left(\dot{\rho}_{(1)} - \frac{1}{2} \rho_{(1)}^2 C'' - \rho_{(1)} \tilde{\cV}'^A d_A \rho_{(1)}
 - \frac{1}{2} \left\| d \rho_{(1)} \right\|^2   \right)  \, . \label{rho2}
\eea
To get this in terms of geometric quantities we substitute in (\ref{rho1}) and apply the zeroth law for SEHs. Thus to second order
\be
\rho^{(\mathcal{E})} \approx - \frac{C}{\kappa_\cV} + \frac{1}{\kappa_\cV^3} \left( C \dot{\kappa}_{\cV} -  \dot{C} \kappa_\cV - \frac{1}{2} C^2 C'' - 2 C \tom^\A d_\A C
+\frac{1}{2} \| dC \|^2 \right) \label{rhoExplicit} \, ,
\ee 
where from  (\ref{d2XC}): 
\be
C'' =- 3 \tom^\A \tom_\A  +  \frac{\tilde{R}}{2} - \sigma^{(\ell)}_{\A\B} \sigma_{(n)}^{\A\B} - \frac{1}{2} \mathcal{R}_{\alpha \beta} \tq^{\alpha \beta} -  \mathcal{R}_{\alpha \beta} \ell^\alpha n^\beta \, . \\
\ee

This is a null surface and in fact this is the only null surface that lives entirely in the regime of the near-horizon approximation. There are other null 
surfaces that pass through the region but this is the only one that remains there. It is clear that in the isolated limit $\rho^{(\mathcal{E})}  \rightarrow 0$
and so one can reasonably argue that if the horizon remains slowly evolving for its entire future and it ultimately asymptotes to isolation, then 
this is the event horizon.

\subsection{Properties of the event horizon candidate}

With our accumulated computational infrastructure it is straightforward to find the geometrical properties of this near-SEH null surface and demonstrate
that it is a slowly evolving null surface. We could work directly from the metric and surface defined by (\ref{rhoExplicit}) but it will be more convenient to 
calculate the required terms as deformations generated by the vector field:
\be
X = - \left(  \frac{C}{\kappa_{\cV}} \right) n \, . 
\ee
First  the induced metric and area/volume element on the $S_v$ that foliate $\mathcal{E}$ are
\bea
\tq^{(\mathcal{E})}_{\A \B} & \approx & \tq_{\A \B} -  \left( \frac{C}{\kappa_{\cV}} \right) k^{(n)}_{\A \B}  \; \; \mbox{and} \\
\vS^{(\mathcal{E})} & \approx & \vS - \left( \frac{C}{\kappa_\cV} \right)  \tn \, .
\eea
As would be expected, to leading order these match the corresponding quantities on $H$. 
This is also the case for other geometric properties with the exception of the expansion which is already at sub-leading order. In that case 
\bea
\tl^{(\mathcal{E})} & \approx & \theta_{(\cV)} + \delta_X \theta_{(\cV)}\\
& \approx & - C \tn  + \left(C \tn + \frac{1}{\kappa_{\cV}} \left( \|\sigma_{(\ell)}  \|^2 + G_{ab} \ell^a \ell^b \right) \right) \nn \\
& \approx & \frac{1}{\kappa_{\cV}} \left( \|\sigma_{(\ell)}  \|^2 + G_{ab} \ell^a \ell^b \right) \, .  \nn
\eea
The first line is just the leading order expansion while the transition to the second applies (\ref{explderiv}), (\ref{expnderiv}) and  (\ref{dXC}) and 
properties of slowly evolving horizons to identify and discard higher order terms. The transition to the final line is then obvious. In any case we recover a ``first'' law for $\mathcal{E}$:
\be
\kappa_{\ell}^{(\mathcal{E})} \tl^{(\mathcal{E})} \approx  \|\sigma_{(\ell)}  \|^2 + G_{ab} \ell^a \ell^b \, , 
\ee
where we have also applied (\ref{dXkappa}) to demonstrate that to leading order $\kappa_\ell^{(\mathcal{E})} \approx \kappa_\cV$. This is a 
direct demonstration of how the first law holds on $\mathcal{E}$. However it would also be straightforward to apply the deformations to 
check that this is a slowly evolving null surface and so must obey a first law. 

Note however, that even at leading order the expansion of $\mathcal{E}$ differs from that of $H$. Combining $(\ref{explderiv})+C \times (\ref{expnderiv})$
and applying the slowly evolving conditions it follows that:
\be
\kappa_\cV \theta_{(\cV)} \approx d_B (d^B C - 2 C \tom^B) + \| \sigma_{(\ell)} \|^2 + G_{ab} \ell^a \ell^b  \, . 
\ee
That is
\be
\kappa_\cV \theta_{(\cV)} - \kappa_{\ell}^{(\mathcal{E})} \tl^{(\mathcal{E})} \approx d_B (d^B C - 2 C \tom^B) \, . \label{expansionDif}
\ee
At leading order the expansions differ by a total derivative. 
The origin of this term can be better understood starting from our defining relation: $[\mathscr{V},X]=0$. Expanding this gives
\bea
[\cV + \tilde{\cV}, X ] = 0 & \Longrightarrow & \delta_X (\delta_\cV \vS) = \delta_{\cV} (\delta_X \vS) - \delta_{[X, \tilde{\cV}]} \vS \, . 
\eea
Thus (again applying properties of slowly evolving horizons)
\bea
\delta_X (\vS \theta_{\cV}) \approx  d_B \left(\delta_X \tilde{\cV}^B \right)
\eea
which leads to (\ref{expansionDif}). The total derivative term is the divergence of the induced shift on $\mathcal{E}$. 
If $C$ is constant (or nearly constant) over a slowly rotating surface, the expansions match at leading order. However in general, 
from the point of view of expansions, the rotation and/or non-constant $C$ induces a mismatch in points on $H$ and $S_v$. In some sense
this is an issue of choice of coordinates: if we integrate the expressions over the $S_v$ and so compare rates of change of 
area, then the total derivative integrates out (for closed $S_v$) and they do match.

%

\subsection{Examples}
Event horizon candidates have been seen previously for certain spacetimes. We now compare our general result with those specific ones. 

\subsubsection{Event horizon candidates in Vaidya spacetimes}

We begin with Vaidya\cite{vaidya,Mukund} spacetimes. The Vaidya metric
\be
ds^2 = - \left(1 -  \frac{2m(v)}{r} \right) dv^2 + 2 dv dr + r^2 \left(d \theta^2 + \sin^2 \theta d \phi^2 \right)
\ee
describes a spherically symmetric black hole that is being irradiated by infalling null dust with stress-energy tensor
\be
T_{ab} =  \frac{dm/dv}{4 \pi r^2} [dv]_a [dv]_b \, . 
\ee
 Any non-decreasing mass function will satisfy the energy conditions. 
Outward and inward-oriented null vectors are given by 
\bea
\ell &=& \frac{\partial}{\partial v} + \frac{1}{2} \left( 1- \frac{2m}{r} \right) \frac{\partial}{\partial r} \; \; \mbox{and} \label{LVaidya} \\
n &=&  - \frac{\partial}{\partial r}  \, . 
\eea 
Then for general $r$
\be
\tl = \frac{r-2m}{r^2} \; \; \mbox{and} \; \;  \tn = - \frac{2}{r}  \, . 
\ee
There is a FOTH located at $r = 2 m$. 

We now specialize to the case of a black hole that transitions from an initial mass $m_1$ to a final mass $m_2 = 2 m_1$. Then $R_H = m_1$ and we scale:
\be
r = R m_1 \; , \; \; v = V m_1  \; \mbox{and} \; \; m = M m_1 \, . 
\ee
The horizon is slowly evolving if $\dot{M} \ll 1$ \cite{billspaper, vaidya} (here the overdot indicates a derivative by $V$). 
Now from (\ref{LVaidya}) outward oriented spherically symmetric null surfaces must be solutions of
\be
\frac{dR}{dV} = \frac{1}{2} \left(1 - \frac{2M(V)}{R} \right)  \, .
\ee
If we assume a hierarchy of derivatives so that $M \gg \dot{M} \gg \ddot{M} \gg \dddot{M} \dots$ then we can search for perturbative solutions 
of the form 
\be
R = 2M \left(1 + \alpha \dot{M} + \beta \dot{M}^2 + \gamma \ddot{M} + \dots  \right) \, .
\ee
Implementing this and switching back to unscaled coordinates, it turns out that to second order there is a null surface at
\be
r^{(\mathcal{E})}  \approx  \; 2m +  8m  \dot{m} + 32m (2 \dot{m}^2 + m \ddot{m})  \, .  \label{EHcandidate}  
\ee
In this expression overdots indicate derivatives with respect to $v$.

This direct calculation matches our more general one. For this scaling of the null vectors it is straightforward to find that on the horizon
\be
\kappa_{\cV} = \frac{1}{4m}  \; \; \mbox{and}  \; \;  C = 2 \dot{m} 
\ee
and since the FOTH at $r=2m$ is spherically symmetric
\be
\tilde{R} = \frac{1}{2 m^2} \, . 
\ee
Further, $-r$ is an affine parameter for the inward moving null geodesics. Setting $\rho = - (r-2m)$ we apply (\ref{rhoExplicit}) and find that 
\bea
\rho_{(1)} & = & - 8 m \dot{m}     \; \; \mbox{and} \\
\rho_{(2)} & = &  - 64 m \dot{m}^2 + 32 m^2 \ddot{m}    \, . \nn
\eea
Converting back into regular $r$ coordinates, this becomes (\ref{EHcandidate}). 

%

\subsubsection{Boost-invariant black brane spacetimes}
Our next example is the five-dimensional black brane spacetime that is the fluid-gravity dual to Bjorken flow\cite{fg1}. 
The spacetime metric takes the form 
\be
ds^2 = - r^2 A(\tilde{\tau}) d \tilde{\tau}^2 + 2 d \tilde{\tau} dr + (1+r \tilde{\tau})^2 e^{b(\tilde{\tau},r)} dz^2 + r^2 e^{c(\tilde{\tau},r)} (dx^2 + dy^2) \, ,
\ee
where the defining functions can be expanded as
\bea
A(\tilde{\tau},r) & = & A_o (v) + \frac{A_1 (v) }{\tilde{\tau}^{2/3}} + \frac{A_2(v)}{\tilde{\tau}^{4/3}} + \dots \\
b(\tilde{\tau},r) & = & b_o (v) + \frac{b_1 (v) }{\tilde{\tau}^{2/3}} + \frac{b_2(v)}{\tilde{\tau}^{4/3}} + \dots \\
c(\tilde{\tau},r) & = & c_o (v) + \frac{c_1 (v) }{\tilde{\tau}^{2/3}} + \frac{c_2(v)}{\tilde{\tau}^{4/3}} + \dots  \, , 
\eea
and $\tau \gg 1$ (so the expansion parameter $\ttau{-2} \ll1$). At the same time $r$ is taken to be sufficiently small so that 
$v = r \ttau{1}$ is always of moderate size. 

Applying the Einstein equations, one can solve order-by-order for the defining function. At lowest order 
\be
A_o (v) = 1 - \frac{\pi^4 \Lambda^4}{v^4} \;  , \; \; b_o (v) = 0 \; \mbox{and} \; \; c_o (v) = 0 
\ee
while at higher orders things become considerably more complicated. Here we will not be concerned with the details of these 
calculations: the results that we need may all simply be read out of \cite{fg1}. In particular it is shown there that: 
%
\bea
r_{EH} &=& \frac{\Lambda}{\ttau{1}} \left\{ \mathsf{r}_o  +  \frac{1}{\Lambda \ttau{2}} \mathsf{r}_1  
+ \frac{1}{\Lambda^2 \ttau{4}} \left( \mathsf{r}_2  + \frac{1}{6 \pi} \right) 
+  \frac{1}{\Lambda^3 \ttau{6}} \left( \mathsf{r}_3 - \frac{29}{432 \pi} - \frac{5}{324 \pi^2} - \frac{17 \log 2}{81 \pi^2}  \right) + \dots \right\} \\
r_{AH} &=& \frac{\Lambda}{\ttau{1}} \left\{ \mathsf{r}_o  +  \frac{1}{\Lambda \ttau{2}} \mathsf{r}_1  
+ \frac{1}{\Lambda^2 \ttau{4}} \left( \mathsf{r}_2  + \frac{1}{9 \pi} \right) 
+  \frac{1}{\Lambda^3 \ttau{6}} \left( \mathsf{r}_3 - \frac{25}{432 \pi} + \frac{1}{81 \pi^2} - \frac{25 \log 2}{162 \pi^2}  \right)  + \dots \right\} 
\eea
where 
\bea
\mathsf{r}_o & = & \pi \\
\mathsf{r}_1 & = & - \frac{1}{2} - \frac{\delta_1}{3} \\
\mathsf{r}_2 & = & \frac{\Lambda \delta_2}{3} - \frac{1}{24} - \frac{\log 2}{18 \pi \Lambda} \\
\mathsf{r}_3 & = & - \frac{1}{7776} +  \frac{\Lambda \delta_3}{3} - \frac{\log (\pi \Lambda)}{18 \pi}+ \frac{C}{18 \pi^2}   + \frac{7 \log^2 2 - 12 \log (\pi \Lambda) }{162 \pi^2} 
\eea
for non-trivial constants $\delta_1$, $\delta_2$ and $\delta_3$. Then a straightforward subtraction gives
\be
r_{EH} - r_{AH} = \frac{1}{\Lambda^2 \ttau{4}} \left(\frac{1}{18 \pi} \right) 
- \frac{1}{\Lambda^3 \ttau{6}} \left( \frac{1}{108 \pi} + \frac{1}{36 \pi^2} + \frac{\log 2}{18 \pi^2}  \right) \, . \label{BoostDist}
\ee 
 
We can compare this with the predicted horizon separation from current calculations. It is also shown in \cite{fg1} that
\be
C = \frac{1}{9 \tau^2} - \frac{1}{\Lambda \ttau{8}} \left(\frac{\log 2}{18 \pi} - \frac{1}{54} \right) + \dots
\ee
and 
\be
\kappa_{\cV} = \frac{1}{\tau^{1/3}} \left(  2 \Lambda \pi - \frac{2}{3 \ttau{2}} + \dots  \right)  \, . 
\ee 
If these are substituted into (\ref{rhoExplicit}) then we should recover (\ref{BoostDist}). The calculation is fairly straightforward. 
Because the metric is vacuum and has planar symmetry, (\ref{rho2}) reduces to
\be
\rho_{(2)} = \frac{1}{\kappa_{\cV} ^3}\left(C \dot{\kappa}_{\cV} - \dot{C} \kappa_{\cV} + \frac{1}{2} 
C^2 \sigma_{\A \B}^{(\ell)} \sigma^{\A \B}_{(n)} \right)   \, . 
\ee
The shear term is of lower order than the others and so may also be neglected ($\| \sigma_{\ell} \| \lessapprox \sqrt{C}$ for 
a slowly evolving horizon). The only complication arises from the fact that both $\kappa_{\cV}$ and $C$ are themselves 
expressed as series. Thus $\rho_{(1)}$ itself contains lower order terms, some of which are of order $\rho_{(2)}$. Once these
are properly accounted for, we recover (\ref{BoostDist}) as expected.

\section{Discussion}
\label{Discuss}

Given geometric data on a (foliated) $n$-dimensional hypersurface $H$, the main work of this paper was to perturbatively reconstruct the 
$(n\!+\!1)$-dimensional spacetime near that surface. We did this in a Gaussian null coordinate system determined by the foliation of $H$. 
To second order in the affine parameter on the inward-oriented null geodesics, the resulting metric was (\ref{metric_expansionII}). At leading
order this is entirely determined by the intrinsic geometry of $H$ while at next order the extrinsic geometry also contributes. However,
 at second and higher orders more knowledge (in the form of the Ricci and Weyl tensors and their derivatives on $H$) is needed. 
These initial calculations were entirely geometric and neither assumed that $H$ was any kind of horizon nor made use any field equations.
They apply to any hypersurface $H$ (that can be foliated by spacelike surfaces) in any spacetime. 

Of course if $H$ is spacelike and the vacuum Einstein equations hold, then it is well-known that there is a good initial value formulation of
general relativity based on the intrinsic and extrinsic metric of $H$ (Section \ref{Sigma}). In that case the intrinsic metric and extrinsic curvature
are sufficient to determine a series expansion of the metric to all orders (in timelike geodesic normal coordinates). This appears to contrast 
with our calculation where, at second order, knowledge of the Weyl tensor is also required. However, a little investigaiton shows that the contradiction 
is only apparent. If $H$ is spacelike ($C \neq 0$) and we assume that the Einstein equations hold, then the required components of the Weyl tensor
can be found from the derivative of the inward-oriented extrinsic curvature ``up'' $H$  via equation (\ref{dkn}):
\bea
 e_A^a n^b e_B^c n^d C_{abcd} 
 & = &  \frac{1}{C} \left( \delta_\cV k^{(n)}_{AB}  + \kappa_\cV k^{(n)}_{ AB} \right)  \label{WeylEq} \\
&& + \left(k^{(n)}_{AC} k^{(n) C}_B 
    - \frac{8 \pi }{(n\!-\!1)} \tq_{AB} {T}_{cd} n^c n^d \right) \nn\\
&& + \frac{1}{C} \left( \frac{1}{2} \tilde{R}_{AB} 
  + \frac{1}{2} \left[ \tn k^{(\ell)}_{AB} + \tl k^{(n)}_{AB} \right]  
  -  2 k^{(n)}_{C(A} k^{(\ell) C}_{B)}  - \frac{1}{2} e_A^a e_B^b \mathcal{R}_{ab} 
  + d_{(A} \tom_{B)} - \tom_A \tom_B \right) 
\, \, . \nn
\eea
Essentially this is a constraint equation on $H$ that determines $e_A^a n^c e_B^b n^d C_{acbd}$ as a function of the intrinsic and extrinsic 
geometry terms. Though we have not done the calculation explicitly, constraints should similarly determine the higher order quantities. 

If $H$ is null the situation is different. In that case $C=0$ and (\ref{WeylEq}) is not well-defined. However this is not surprising since in this
case we would not expect a good initial value formulation: the domain of dependence of a null surface is empty. Physically this is because  
extra information that has travelled ``parallel'' to $H$ can influence spacetime arbitrarily close to $H$. 
As mentioned earlier, to get a good initial value formulation, data must be specified on a pair of intersecting null surfaces \cite{friedrich}. 
Thus apart from on an isolated 
horizon one would also need to specify data on, for example, an outgoing past-directed null cone originating from some $S_{v_o}$.
Specifically in vacuum one would need $\tq_{AB}$, $k^{(\ell)}_{AB}$, $k^{(n)}_{AB}$ and $\tom_A$ on $S_v$ along with 
 $e_A^a \ell^c e_B^b \ell^d C_{acbd}$ on $H$ (this must vanish for an isolated horizon) 
 and  $e_A^a n^c e_B^b n^d C_{acbd}$ on the past-oriented null-cone \cite{badri,lewandowski}. 
This would be sufficient to specify the spacetime for $v>v_o$ and $\rho < 0$ (at least while the coordinates are well-defined).

We considered two applications of these results. The first was a straightforward reconstruction of the spacetime near an extremal isolated 
horizon. At leading order this can be done without any need for off-horizon information and the result is a near-horizon spacetime in the
sense of \cite{hariOld,hariClass}. In fact, those near-horizon spacetimes are actually exact solutions of the Einstein equations: in this case, 
throwing away sub-leading corrections turned an approximate solution into an exact one! The details of exactly why this happens deserve
further consideration but we leave this small puzzle for later investigation. 

The second example was more involved and  we demonstrated the existence of an event horizon 
candidate which hugs any slowly evolving horizon. This general result was foreshadowed by similar results in the case of Vaidya and 
near-equilibrium black brane spacetimes, however we believe that this is the first time that it has been demonstrated in full generality. It should
be emphasized that this tentative identification does not violate the teleological nature of true event horizons: in order to identify this 
SEH-hugging null surface as a true event horizon we must assume that the trapping horizon remains slowly evolving for the rest of eternity
and that it ultimately settles down (or at least asymptotes) to equilibrium. If both of these are true, then we have identified a null surface that
asymptotes to that equilibrium state and so is the event horizon. 

Of course in all of this work we have assumed that the series expansion of the metric converges. We expect this to be the case. Higher order 
derivatives of our geometric quantities will also depend on surface quantities and their derivatives and we do not expect them to blow up. 
However, this is not a completely trivial result. Our first attempt to demonstrate the existence of an event horizon 
candidate near slowly evolving horizons used the standard $(n\!+\!1)$-formalism from Section  \ref{Sigma} to expand spacetime around the 
(spacelike) dynamical slowly evolving horizon \cite{SEHreview}. At leading order the results of that work essentially matched those 
found here. However subsequent calculations have demonstrated that the second order ``corrections'' in the standard expansion are 
actually proportional to $1/\sqrt{C}$: in timelike 
geodesic coordinates the series analogous to (\ref{HorLoc}) \emph{does not} converge. In retrospect this is not so surprising. Relative to our 
Gaussian normal coordinate system, the inward oriented timelike normal to a slowly evolving horizon is 
\be
\hat{\tau}^a = \frac{1}{\sqrt{2 C}} \left( \frac{\partial}{\partial v} \right)^a + \sqrt{\frac{C}{2}}  \left( \frac{\partial}{\partial \rho} \right)^a \, ,
\ee
and so time derivatives relative to this vector pick up factors of $(2C)^{-1/2}$. Intuitively small changes in proper time along the normal 
geodesics can correspond to large changes in the horizon coordinate $v$ and therefore significant changes in the usual geometric quantities. 

In conclusion we briefly consider future applications of this formalism. It is easy to see that ``just outside'' the event 
horizon candidate there will a timelike surface that similarly hugs any slowly evolving horizon. Such a surface could be treated as a
\emph{stretched horizon} from membrane paradigm \cite{membrane}. Via this link much of the membrane paradigm formalism 
will be translatable into the language of slowly evolving horizons and their accompanying surfaces (and vice versa). We expect
that new astrophysical insights may follow from this cross-fertilization with obvious targets of study including black hole ring-downs\cite{Berti:2007fi}, 
spin-flips\cite{Campanelli:2006fy}
post-merger recoils\cite{Baker:2006vn} and anti-kicks\cite{Jaramillo:2011re,Pollney:2007ss}. It will also be possible to write many of the membrane paradigm results in a form that will apply to general 
black holes and branes in general dimensions. Connections with blackfolds \cite{blackfolds} are an obvious target of study.

%
%

\appendix

\section*{Acknowledgements}
I.B. was supported by the Natural Sciences and Engineering Research Council of Canada. Many of the calculations and parts of the writing 
of this paper were done while he was on sabbatical leave at the University of Barcelona and he would like to thank  
the Departament de F\'{i}sica Fonamental for their hospitality during that time. He also thanks 
Stephen Fairhurst, Hari Kunduri, Aghil Alaee  and David Tian for useful conversations. 

\section{Kerr-Newman in EF-normal coordinates}
\label{KerrEx}

In this appendix we demonstrate the construction of the Gaussian null coordinate system for a Kerr-Newman horizon. To begin, 
recall that in standard Eddington-Finkelstein coordinates the Kerr-Newman spacetime takes the form:
\bea
ds^2 &=& - \left(1 - \frac{\Delta-\chi}{\Sigma} \right) dv^2 + 2 dv dr - \frac{2a (\chi - \Delta) \sin^2 \!\theta}{\Sigma} dv d\phi \label{KerrEF}\\
& & - 2 a \sin^2 \!\theta dr d\phi + \Sigma d\theta^2 + \frac{\sin^2 \!\theta (\chi^2 - a^2 \Delta \sin^2 \!\theta)}{\Sigma} d \phi^2 \nn
\eea
where $\Delta = r^2 - 2mr +a^2 +Q^2$, $\chi = r^2 + a^2$  and $\Sigma = r^2 + a^2 \cos^2 \! \theta$.  
The isolated horizon $H$ is at $r_o$, the larger root of $\Delta(r)$ and we consider the foliation $S_v$ of surfaces of constant $v$. 
Note that though  based on ingoing null geodesics there are two things that distinguish these standard coordinates from Gaussian null
coordinates:
\begin{enumerate}
\item On $H$, the $dv d\phi$-term is non-vanishing. That is,  $\frac{\partial}{\partial v}$ is \emph{not} orthogonal to 
the foliation surfaces of constant $v$ and as such it is not a $\mathcal{V}^a$ candidate. 
\item The coordinate $r$ is an affine parameter for a family of ingoing future oriented null geodesics \emph{but} that family is not orthogonal to the $S_v$ 
(as evidenced by the non-zero $dr d \phi$ term). 
\end{enumerate} 

The first of these difficulties is easily resolved by the coordinate transformation 
\be\phi = \varphi + \left( \frac{a}{\chi_o} \right) v
\ee
with $\chi_o = r_o^2 + a^2$ which ``unwinds''
the $S_v$ on the horizon. The metric becomes:
\bea
ds^2 &=& - \left(\frac{\Sigma_o^2 \Delta - a^2 \sin^2 \! \theta (r-r_o)^2}{\Sigma \chi_o^2} \right) dv^2 + \frac{2\Sigma_o}{\chi_o^2} dv dr 
 + \frac{2 a \sin^2 \! \theta}{\Sigma (r_o^2 + a^2)} \left(\Delta \Sigma_o + \chi (r^2-r_o^2) \right)dv d\phi \label{KerrEF}\\
& & - 2 a \sin^2 \!\theta dr d\phi + \Sigma d\theta^2 + \frac{\sin^2 \!\theta (\chi^2 - a^2 \Delta \sin^2 \!\theta)}{\Sigma} d \phi^2 \nn
\eea
where $\Sigma_o = r^2 + a^2 \cos^2 \! \theta$. Then the full (null) three-metric on $H$ is
\be
dS^2 = \Sigma_o d\theta^2 + \frac{\sin^2 \! \theta \chi_o^2}{\Sigma_o} d \phi^2
\ee
which is now of the expected form. 

The second difficulty is resolved by rewriting the metric relative to the correct set of geodesics. Relative to the current coordinate system a suitable pair of 
(cross-normalized) null normals is
\be
\ell = \frac{\partial}{\partial v} \; \; \; \mbox{and} \; \; \; n = - \left(\frac{a^2 \sin^2 \! \theta}{2 \Sigma_o} \right) \frac{\partial}{\partial v} 
- \left( \frac{r_o^2 + a^2}{\Sigma_o} \right) \frac{\partial}{\partial r} -  \left( \frac{a}{r_o^2 + a^2} \right) \frac{\partial}{\partial \phi}  \, . 
\ee
Keep in mind that these are defined only on the horizon. Now consider the family of null geodesics that crosses $H$ with $n$ as its tangent vector field.
We identify them by the point $(v, \theta, \varphi)$ where they cross $H$ and parameterize them with affine parameter $\rho$ so that $\rho=0$ on $H$ 
and increases inwards. We perturbatively construct the geodesics up to third order in $\rho$:
\bea
X_{(v, \theta, \varphi)}^\alpha (\rho) \approx \left. X^\alpha \right|_{\rho = 0} + \rho  \left. \frac{dX^\alpha}{d\rho} \right|_{\rho=0}  
+ \frac{\rho^2}{2}  \left. \frac{d^2 X^\alpha}{d \rho^2} \right|_{\rho = 0}  
+ \frac{\rho^3}{6}  \left. \frac{d^3 X^\alpha}{d \rho^3} \right|_{\rho = 0} \, . \label{Xdef}
\eea
On the right-hand side of this equation and henceforth, the labelling subscript $(v, \theta, \varphi)$ is omitted but understood. 

The first two coefficients are trivial:
\be
\left. X^\alpha \right|_H = [v, r_o , \theta, \varphi]  
\ee
and 
\be
\left. \frac{d X}{d \rho}^\alpha  \right|_H = n^\alpha 
= \left[- \frac{a^2 \sin^2 \! \theta}{2 \Sigma_o}, - \frac{r_o^2 + a^2}{\Sigma_o}, 0,  -   \frac{a}{r_o^2 + a^2} \right] \, . 
\ee
The next two follow from the geodesic equation: 
\be
n^\beta \nabla_\beta n^\alpha = 0 \;  \; \Rightarrow \; \;  \left. \frac{d^2 X^\alpha}{d \rho^2} \right|_H = 
-  \Gamma_{\beta \gamma}^\alpha n^\beta n^\gamma
\ee
and 
\be
n^\gamma \nabla_\gamma \left(n^\beta \nabla_\beta n^\alpha  \right) = 0 \; \; \Rightarrow \; \; 
 \left. \frac{d^3 X^\alpha}{d \rho^3} \right|_H  =  \left(-  \partial_\beta \Gamma_{\phantom{\alpha} \gamma \delta}^\alpha 
+ 2 \Gamma^\alpha_{\phantom{\alpha} \beta \epsilon} \Gamma^\epsilon_{\phantom{\alpha} \gamma \delta}  \right)  n^\beta n^\gamma n^\delta \,. 
\ee
The right-hand sides follow from expanding the left-hand side and making appropriate substitutions from the earlier derivatives. The 
Christoffel symbols $\Gamma^\alpha_{\phantom{\alpha} \beta \gamma}$ and their derivatives 
$\partial_\beta \Gamma_{\phantom{\alpha} \gamma \delta}^\alpha$ only need to be evaluated on $H$. 

Once these quantities are calculated (\ref{Xdef}) defines a transformation from $(v,r,\theta,\varphi)$ to $(v,\rho,\theta,\varphi)$
coordinates. Then the second order expansion of the metric is:
\be
g_{\alpha \beta} \approx g^{(0)}_{\alpha \beta} + \rho g^{(1)}_{\alpha \beta} + \frac{\rho^2}{2} g^{(2)}_{\alpha \beta} 
\ee
where the zeroth order components are
\bea
g_{\theta \theta}^{(0)} & = & \Sigma_o \\
g_{\varphi \varphi}^{(0)} & = &\left( \frac{ \chi_o^2 }{\Sigma_o} \right) \sin^2 \! \theta  \nn  \, ,
\eea
the first order corrections are found to be 
\bea
g_{vv}^{(1)} & = & \frac{\Delta'}{\chi_o} \\
g_{v\theta}^{(1)} & = & \nn -\frac{2a^2\sin\!\theta\cos\!\theta}{\Sigma_o}\\
g_{v\varphi}^{(1)} & = & \nn - \left( \frac{a \sin^2 \! \theta}{ \Sigma_o} \right) \Delta'- \left( \frac{2a r_o \chi_o \sin^2 \! \theta}{\Sigma_o^2} \right)\\
g_{\theta \theta}^{(1)} & = & - \frac{2r_o \chi_o}{\Sigma_o}  \nn \\
g_{\theta \varphi}^{(1)} & = & \nn  \frac{2a^3 \chi_o \sin^3\!\theta \cos\!\theta}{\Sigma_o^2}  \\
g_{\varphi \varphi}^{(1)} & = & \left( \frac{a^2 \chi_o \sin^4 \! \theta}{\Sigma_o^2} \right)  \Delta'  
- \left( \frac{2r_o \chi_o^2 \sin^2 \!\theta (\Sigma_o-a^2 \sin^2 \!\theta)}{\Sigma_o^3} \right) \nn 
\eea
and the second order corrections are (the somewhat complicated):
\bea
g_{vv}^{(2)} & = & \left( \frac{a^2  \sin^2 \!\theta}{4 \chi_o^2 \Sigma_o  } \right) (\Delta')^2 + \left( \frac{ r_o(2\chi_o +a^2 \sin^2 \! \theta)}{\chi_o \Sigma_o^2 } \right) \Delta' -\left( \frac{\Sigma_o^2-4a^2r_o^2 \sin^2 \! \theta}{\Sigma_o^3} \right)\\
g_{v\theta}^{(2)} & = & - \left(\frac{a^2 \sin \! \theta \cos \! \theta (3 \chi_o^2 + a^2 \sin^2 \! \theta)}{2 \chi_o \Sigma_o^2 } \right) \Delta' 
- \left( \frac{2 a^2 r_o \sin \! \theta \cos \! \theta (\chi_o^2 + a^2 \sin^2 \! \theta)}{\Sigma_o^2} \right) \nn \\
g_{v\varphi}^{(2)} & = & - \left(\frac{a^3 \sin^4 \! \theta}{4 \chi_o \Sigma_o^2} \right) (\Delta')^2 
- \left( \frac{a r_o \sin^2 \! \theta (4 \chi_o + 3 a^2 \sin^2 \! \theta ) }{\Sigma_o^3} \right) \Delta' 
+ \left(\frac{a \chi_o \sin^2 \! \theta (2 \Sigma_o^2 - a^2 \sin^2 \! \theta \left(5 r_o^2 - a^2 \cos^2 \! \theta) \right)}{\Sigma_o^4} \right) \nn \\
g_{\theta \theta}^{(2)} & = & - \left( \frac{a^2 r_o \sin^2 \! \theta}{2 \Sigma_o^2} \right) \Delta'
+ \left(  \frac{r_o^6 + (1 + \cos^2 \! \theta) a^2 r_o^4 - (5 \cos^4 \! \theta + 7 \cos^2 \! \theta - 1) a^4 r_o^2 + \sin^2 \! \theta \cos^2 \! \theta (\cos^2 \! \theta - 5) a^6}{\Sigma_o^3}  \right) \nn \\
g_{\theta \varphi}^{(2)} & = & \left( \frac{a^3 (4 \chi_o - \Sigma_o) \sin^3 \! \theta \cos \! \theta }{2 \Sigma_o^3} \right) \Delta' 
+ \left( \frac{a^3 r_o \chi_o \sin^3 \! \theta \cos \! \theta (\Sigma_o + 6 a^2 \sin^2 \! \theta )}{\Sigma_o^4} \right) \nn \\
g_{\varphi \varphi}^{(2)} & = & \left( \frac{a^4 \sin^6 \! \theta}{4 \Sigma_o^3} \right) (\Delta')^2 + \left(\frac{a^2 r_o \chi_o  \sin^4 \! \theta (3 \chi_o + 4 a^2 \sin^2 \theta) }{\Sigma_o^4} \right) \Delta' \nn \\
& &  + \left( \frac{ \chi_o^2 \sin^2 \! \theta\left(r_o^6 + (5 \cos^2 \! \theta - 1) a^2 r_o^4 + (11 \cos^4 \! \theta - 15 \cos^2 \! \theta + 5) a^4 r_o^2 + \sin^2 \theta \cos^2 \! \theta (1 + \cos^2 \! \theta)a^6 \right) }{\Sigma_o^5} \right)  \nn  \,. 
\eea

These same expressions are also found when calculating directly from (\ref{metric_expansion}). This, of course, isn't a surprise but it 
does provide a reassuring cross-check on potential typographical errors.

%

\end{document}